\documentclass[twocolumn,aps,prb,superscriptaddress,nofootinbib,float,longbibliography]{revtex4-1}
\usepackage[colorlinks,bookmarks=false,citecolor=blue,linkcolor=red,urlcolor=blue]{hyperref}
\usepackage{verbatim}
\usepackage{amsmath,amssymb,bm,braket,float,mathtools}
\usepackage{graphicx,xfrac,appendix}
\usepackage[english]{babel}
\makeatletter
\usepackage{epstopdf}
\makeatother

\begin{document}
\title{Power law hopping of single particles in one-dimensional non-Hermitian quasicrystals}
\author{Dechi Peng}
\affiliation{Department of Physics, Zhejiang Normal University, Jinhua 321004, China}
\author{Shujie Cheng}
\thanks{chensj@zjnu.edu.cn}
\affiliation{Department of Physics, Zhejiang Normal University, Jinhua 321004, China}
\author{Gao Xianlong}
\thanks{gaoxl@zjnu.edu.cn}
\affiliation{Department of Physics, Zhejiang Normal University, Jinhua 321004, China}
\date{\today}

\begin{abstract}
In this paper, a non-Hermitian Aubry-André-Harper  model with power-law hoppings ($1/s^{a}$) and quasiperiodic  parameter $\beta$ is studied,
where $a$ is the power-law index,  $s$ is the hopping distance, and $\beta$ is a member of the metallic mean family. We find that under the weak
non-Hermitian effect, there preserves $P_{\ell=1,2,3,4}$ regimes  where the fraction of ergodic
eigenstates is $\beta$-dependent as $\beta^{\ell}$L ($L$ is the system size) similar to those in the Hermitian case. However, $P_{\ell}$ regimes are ruined by the strong non-Hermitian effect.
Moreover, by analyzing the fractal dimension, we find that there are two types of edges aroused by the power-law index $a$ in the
single-particle spectrum, i.e., an ergodic-to-multifractal  edge for the long-range hopping case ($a<1$), and an ergodic-to-localized  edge
for the short-range hopping case ($a>1$). Meanwhile, the existence of these two types of edges is found to be robust against the non-Hermitian
effect. By employing the Simon-Spence theory, we analyzed the absence of the localized states for $a<1$. For the short-range hopping case,
with the Avila's global theory and the Sarnak method, we consider a specific example with $a=2$ to reveal the presence of the intermediate
phase and to analytically locate the intermediate regime and the ergodic-to-multifractal  edge, which are self-consistent with the numerically results.
\end{abstract}
\maketitle

\section{introduction}
In 1958, P. W. Anderson pointed out that free particles will present localized behaviors due to  random disorders.
The absence of diffusion is known as  Anderson localization  \cite{PhysRev.109.1492}.
The scaling theory shows\cite{PhysRevLett.42.673,PhysRevLett.100.013906,PhysRevLett.105.163905} that systems change from fully ergodic
phase to fully localized phase with the arbitrarily weak disorder in one and two dimensional (1D and 2D) Anderson model. However, for 3D case,
an energy threshold, i.e., the mobility edge, appears in the single-particle spectrum and separates the ergodic eigenstates from the localized eigenstates.
Beyond the Anderson-like model, the mobility edge appears in a class of generalized Aubry-André-Harper (AAH) models as well. It is known to us
that there is no any mobility edge in the standard  AAH model \cite{aubry1980analyticity,1982JETP...56..612S,wilkinson1984critical}, but the mobility edges can be induced by breaking the self-duality \cite{PhysRevLett.104.070601,PhysRevB.106.144208,PhysRevLett.123.025301,PhysRevB.103.075124,PhysRevB.102.024205},
such as introducing the next-nearest-neighbor hoppings \cite{PhysRevB.83.075105}, the exponentially long-range hoppings \cite{PhysRevLett.104.070601,PhysRevB.103.134208}, the off-diagonal incommensurate hoppings \cite{PhysRevLett.127.116801,PhysRevB.106.144208,PhysRevB.104.085401,PhysRevB.103.134208,PhysRevB.103.014203}, the power-law hoppings \cite{PhysRevLett.123.025301,PhysRevB.103.075124,PhysRevB.83.075105,PhysRevB.100.174201}, the slow-varying potentials \cite{Cheng_2022,PhysRevB.105.104201}, and the generalized incommensurate potentials\cite{Farchioni1993IncommensuratePA,art,YUCE20142024,PhysRevA.95.062118,PhysRevB.100.054301,PhysRevLett.122.237601,PhysRevResearch.2.033052,PhysRevB.103.014203}.
The studies on single-particle mobility edge\cite{PhysRevA.94.033615,PhysRevA.98.013635,PhysRevB.90.054303,Roy_2021,PhysRevB.101.064203,PhysRevB.96.085119,weidemann2022topological,PhysRevB.103.144202,PhysRevB.102.024205,PhysRevA.105.063327,PhysRevB.105.205402} help us understand the roles that mobility edge plays on the thermalization and many-body
localization in interacting quasidisordered extensions\cite{unknown,Bordia_2017,PhysRevLett.123.090603}. 

Recently, there are growing interests in studying the mobility edges in a class of generalized AAH model
with power-law hoppings \cite{PhysRevLett.123.025301,PhysRevB.83.075105,PhysRevB.103.075124,PhysRevLett.120.160404}, which can be induced by power-law interactions \cite{PhysRevB.83.075105,PhysRevLett.120.160404}. Deng et al. found that
when the power-law index $a<1$, there are ergodic-to-multifractal (EM) edges in the intermediate regimes,
and when $a>1$, there are ergodic-to-localized (EL) edges\cite{PhysRevLett.123.025301}. Particularly, the intermediate regimes is subdivided into $P_{\ell}$ regimes,
where the fraction of the ergodic states are of $\beta^{\ell}_{g}$L ($\beta$ is a quasiperiodic parameter measuring the member of the metallic mean family, $L$ is the system size, and {$\ell=1, 2, 3, 4$}). Roy and Sharma discussed the
influence of the metallic mean family on the intermediate regime, and a generalized phase diagram based on the irrational
Diophantine numbers and their sequences are charted out \cite{PhysRevB.103.075124}. Xu et al.
studied the non-Hermtian effect on the  power-law hopping system \cite{PhysRevB.104.224204}, and found that  the aforementioned $P_{\ell}$ regimes
are destroyed by the non-Hermitian effect and the EM and EL edges
are independent of the quasiperiodic parameter $\beta$. Besides, the localization transition points and the exact
expression of the EL edge are derived, which are self-consistent with numerical results. In this work, we are motivated
to study whether the $\beta$-dependent $P_{\ell}$ regimes are robust against
the non-Hermitian transition effect.
In addition, we will try to understand the absence of the localized states in the long-range hopping regime and analytically obtain
the EL edges in the short-range hopping regime (such as $a=2$).

The organization of the paper is as follows. In Sec. \ref{S2}, we describe the Hamiltonian of the
non-Hermitian AA model with power-law hopping, and introduce the metallic mean family.
In Sec. \ref{S3}, we study the localization properties under the weak non-Hermitian effect.
In Sec. \ref{S4}, we study the localization properties under the strong non-Hermitian effect.
For $h=0.8$, we studied  the relationship between the localization transition of the eigenstates and the breaking of the $PT$ symmetry in the case of both $a<1$ and $a>1$ in  Sec. \ref{S5}. We summarized in Sec. \ref{S6}.

\section{MODEL AND HAMILTONIAN}\label{S2}
A one-dimensional non-Hermitian AA model that we considered consists of  power-law
hoppings and complex on-site potentials, and it reads	
\begin{equation}\label{eq1}
	H =-J \sum_{j,s} \frac{1}{s^{a}}\left(c_{j}^{\dagger} c_{j+s}+H.c\right)
	  +\sum_{j} \Delta_{j} c_{j}^{\dagger} c_{j},
  \end{equation}
where $J$ is set as the unit of energy, $a$ is the power-law index, $J/s^{a}$ is the power-law hopping strength between site $j$ and site $j+s$,
and $\Delta_{j}= \Delta \cos (2 \pi\beta j +ik)$ denotes the non-Hermitian on-site potential. The non-Hermitian effect
is introduced by an imaginary term $ik$. When $k=0$, the model goes back to the Hermtian case \cite{PhysRevLett.123.025301,PhysRevB.103.075124},
where EM edges are uncovered. $\Delta_{j}$ satisfies the relation $\Delta_{-j}=\Delta^{*}_{j}$,
therefore the Hamiltonian $H$ is $PT$-symmetric \cite{PhysRevLett.122.237601,PhysRevB.104.224204}. $\beta$ is chosen at the  metallic mean family, which can be derived from a generalized $u$-Fibonacci
recurrence relation $F_{v+1}=uF_v+F_{v-1}$ with $F_0=0$ and $F_1=1$. The golden mean $\beta=\beta_{g}$ is obtained by the limit
$\beta_{g}=\lim _{v \rightarrow \infty} F_{v-1} / F_v$ when $u=1$. Besides, this recurrence can yield another metallic mean,
such as the silver mean $\beta=\beta_s=\sqrt{2}-1$  when $u=2$ and the bronze mean $\beta=\beta_b=(\sqrt{13}-3)/2$ when $u=3$.
$\beta_{g}$  and $\beta_{s}$ will be used in following numerical calculations.

With the basis $\left|\psi_{n}\right\rangle=\sum_{j} \phi^{n}_{j}|j\rangle=\sum_{j} \phi^{n}_{j} c_{j}^{\dagger}|0\rangle$,
we obtain the following eigenfunction:
\begin{equation}\label{eigen_1}
-J\sum_{s}\frac{1}{s^{a}}\left(\phi^{n}_{j-s}+\phi^{n}_{j+s}\right)+\Delta_{j} \phi^{n}_{j}=E_{n} \phi^{n}_{j},
\end{equation}
where $\phi^{n}_{j}$ is the amplitude at the $j$th site of the $n$th wave function,
and $E_{n}$ is the corresponding eigenenergy. Here the eigenenergy levels with ascending order are sorted according to the real part of $E_{n}$.
		



\section{Localization properties under weak non-Hermitian effect}\label{S3}

\begin{figure}[t]
	\begin{minipage}[h]{1.0\linewidth}
	\centering
	\includegraphics[width=1.0\textwidth]{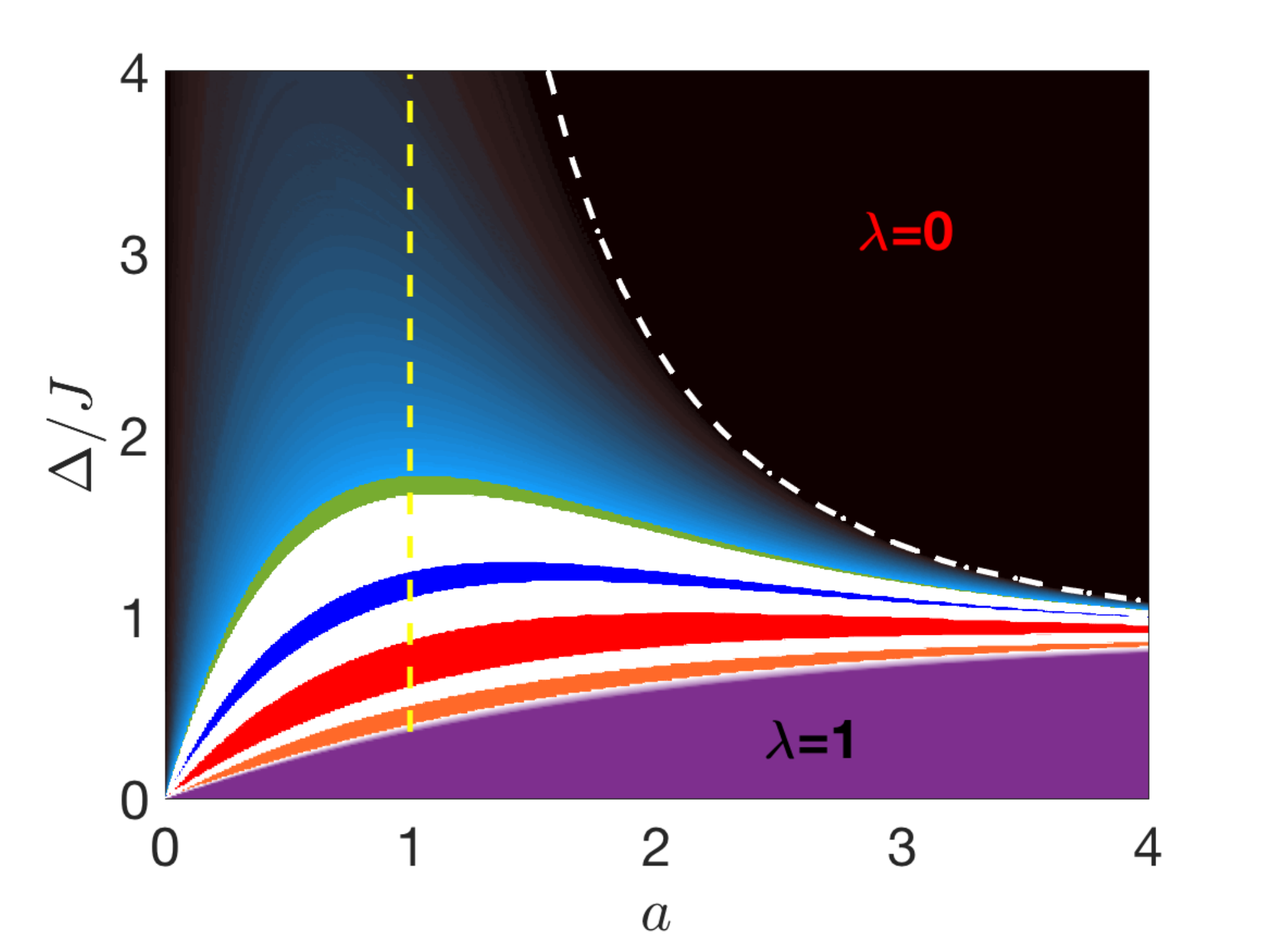}
	\renewcommand\figurename{Fig}
\end{minipage}
	
\renewcommand\figurename{Fig}
\caption{ The phase diagram of the non-Hermitian AAH model with  power-law hopping index $a$ and the strength of the complex
potential $\Delta$  for $\beta_g =610/987$, $k=0.8$ and the system size $L = 987$.  In addition to ergodic (purple regime) and
localized (black regime) phases with  the fraction of ergodic eigenstates $\lambda = 1$ and $\lambda = 0$ respectively, there is
an intermediate phase with $0<\lambda<1$. In particular, in the intermediate phase,
 there are four $P_{\ell=1,2,3,4}$ regimes with fractions  $\lambda=\beta_g$ (orange regime, marked by $P_{1}$),
$\beta_g^2$ (red regime, marked by $P_{2}$), $\beta_g^3$ (blue regime, marked by $P_{3}$),
and $\beta_g^4$ (green regime, marked by $P_{4}$).
 }\label{f1}
\end{figure}

As mentioned before, the parameter $k$ dominates the non-Hermitian effect. When $k$ is mall,  the non-Hermitian effect is weak,
whereas it is strong when $k$ is large. In this section, we mainly study the weak non-Hermitian case with  $k=0.8$ and $\beta=\beta_{g}$.
The phase diagram of the model in Eq. (\ref{f1}) with $\beta=\beta_{g}$ has been presented in Fig.~\ref{f1}.
We find that the non-Hermitian system preserves  similar features as the Hermitian one. For $a\gg 1$, we recover the non-Hermitian
AAH model \cite{PhysRevLett.122.237601} with nearest-neighbor hoppings,
and therefore all eigenstates are either ergodic (the purple regime with  the fraction of ergodic eigenstates $\lambda=1$) for $\Delta<2e^{-k}$ or localized (black regime with the fraction of ergodic eigenstates $\lambda=0$) for $\Delta>2e^{-k}$.
As can be seen from the phase diagram  in addition to ergodic (purple regime) and localized (black regime) phases with
the fraction of ergodic eigenstates $\lambda = 1$ and $\lambda = 0$ respectively, there is an intermediate phase with $0<\lambda<1 $.
In particular, in the intermediate phase, there are four $P_{\ell=1,2,3,4}$ regimes where the lowest $\beta^{\ell}_{g}$L eigenstates
are ergodic with fractions  $\lambda=\beta_g$ (orange regime, marked by $P_{1}$),
$\beta_g^2$ (red regime, marked by $P_{2}$), $\beta_g^3$ (blue regime, marked by $P_{3}$),
and $\beta_g^4$ (green regime, marked by $P_{4}$).
Compared with the Hermitian cases \cite{PhysRevLett.123.025301}, the remarkable differences are reflected in that the four regimes are
suppressed by the non-Hermitian effect, and are separated by the normal intermediate regimes, where the fraction of ergodic states are $\beta$-independent.  Meanwhile, the original $P_{\ell>4}$ regimes no more exist.  In the following, we will clarify the similarities and differences between the $P_{\ell}$ regimes and the normal intermediate regime by investigating the fractal dimension.

\begin{figure}[htp]
		\centering
		\includegraphics[width=0.50\textwidth]{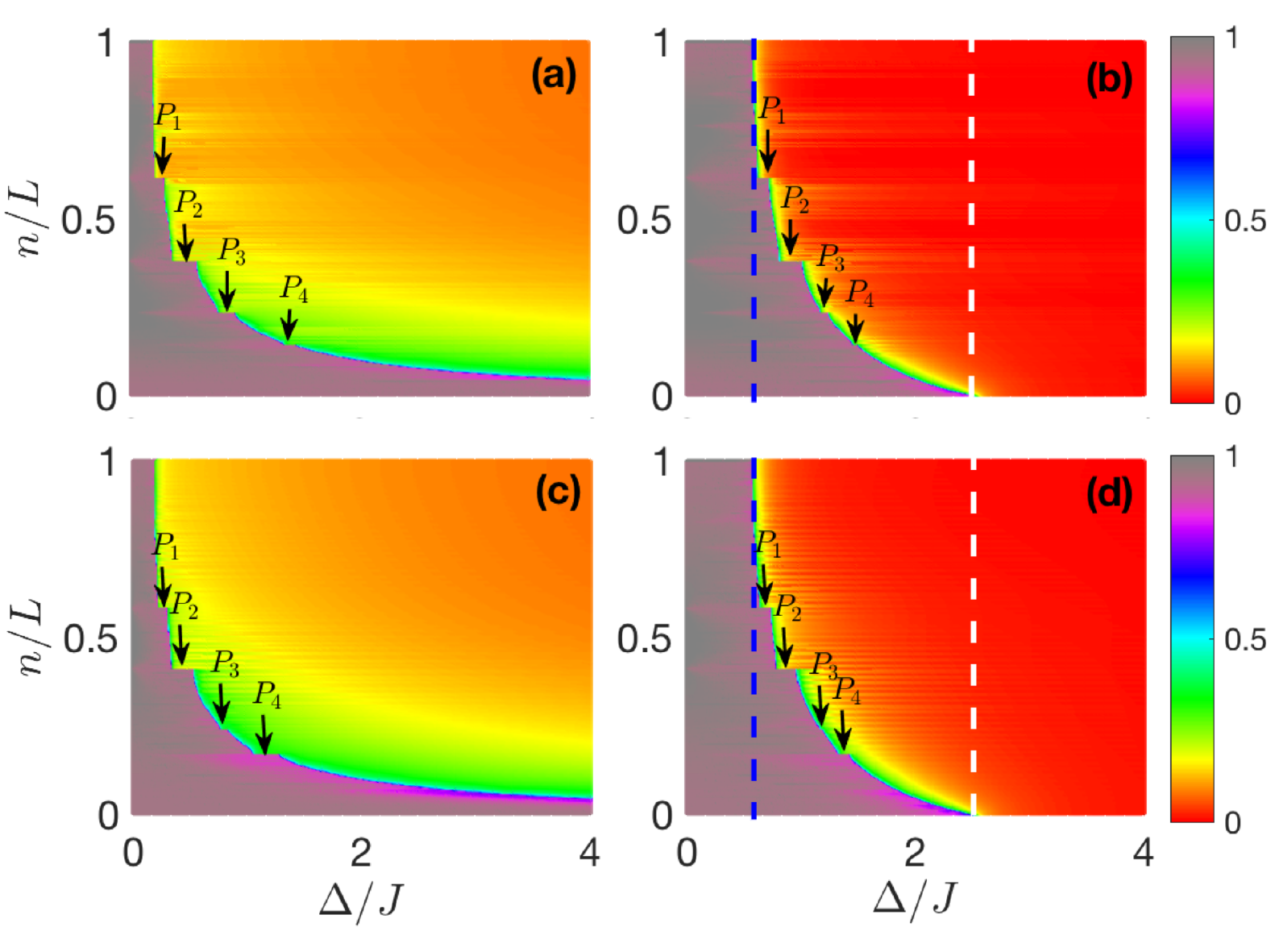}
		\renewcommand\figurename{Fig}
	\caption{  Fractal dimension $D_2 $ (shown in color) of different eigenstates as a function of $\Delta$ with $k=0.8$, $L=2584$, $d=4$,
	and $\beta_g =1597/2584$ for (a) $a=0.5$  and (b) $a=2$, and with $k=0.8$, $L=2378$, $d=2$,
	and $\beta_s =985/2378$ for (c) $a=0.5$  and (d) $a=2$. In (b) and (d), the blue and white dashed lines correspond to
	$\Delta_{c 1} \approx 0.6J$ and $\Delta_{c 2} \approx 2.4J$, respectively.}\label{f2}
\end{figure}

The fractal dimension $D_{f}$ is defined based on the box counting procedure \cite{PhysRevLett.62.1327,doi:10.1142/S021797929400049X,RevModPhys.67.357,PhysRevB.68.184206}
and is expressed as
\begin{equation}
D_f=\lim _{L_{d} \rightarrow \infty } \frac{1}{1-f} \frac{\ln \sum_{m=1}^{L_{d}}\left(\mathcal{I}_m\right)^f}{\ln L_{d}},
\end{equation}
where $L_d=L / d$ is the number of the box with $L$ being the system size and $d$ being the box counting index,
$f$ is the scale index, and $\mathcal{I}_m=\sum_{j \in m}\left|\psi_n(j)\right|^2$ corresponds to the probability of
detecting inside the $m$th box for the $n$th normalized eigenstate $|\psi_n(j)\rangle $.
Without loss of generality, we study the fractal dimension $D_{2}$. Considering the system size $L=2584$,
and the box counting index $d=4$,  as well as the golden mean $\beta_{g}=1597/2584$, we plot  $D_{2}$ of full eigenstates
as a function of  the strength of the complex potential $\Delta$ for $a=0.5$ (long-range hopping) in Fig.~\ref{f2}(a) and for $a=2.0$ (short-range hopping) in
Fig.~\ref{f2}(b), respectively. It is readily seen that, in the $P_{\ell=1,2,3,4}$ regimes, two types of edges present a step-wise dependence
on $\Delta$, equaling to $\lambda=\beta^{\ell}_{g}$. Out of the four regimes, $\lambda$ smoothly changes as $\Delta$ increases.

In fact, similar phenomena appear in the $\beta=\beta_{s}$ case as well. For  systems size $L=2378$ and different box counting index $d=2$, as well as the silver mean  $\beta_{s}=985/2378$, we plot $D_{2}$ as a function of $\Delta$ for $a=0.5$ (long-range hopping) in Fig.~\ref{f2}(c) and for $a=2.0$ (short-range hopping) in Fig.~\ref{f2}(d), respectively. Compared to Figs.~\ref{f2}(a) and \ref{f2}(b), the two types of edges display  a  different  step-wise dependence on the $\Delta$ in the $P_{\ell=1,2,3,4}$ regimes ($\lambda=\beta_{s}+\beta^{2}_{s} $ $(P_{1})$, $\beta_{s}$ $ (P_{2})$, $\beta^{2}_{s}+\beta^{3}_{s}$ $(P_{3})$, $2\beta^{3}_{s}+\beta^{4}_{s}$ $(P_{4})$, respectively) and a same smooth changing of  $\lambda$ out of the $P_{\ell}$ regimes still exist. It implies that the  step-wise dependence on the $\Delta$ in the $P_{\ell=1,2,3,4}$ regimes depends on the quasiperiodic  parameter $\beta$.

\begin{figure}[htp]
	
	\centering
	\includegraphics[width=9.4cm]{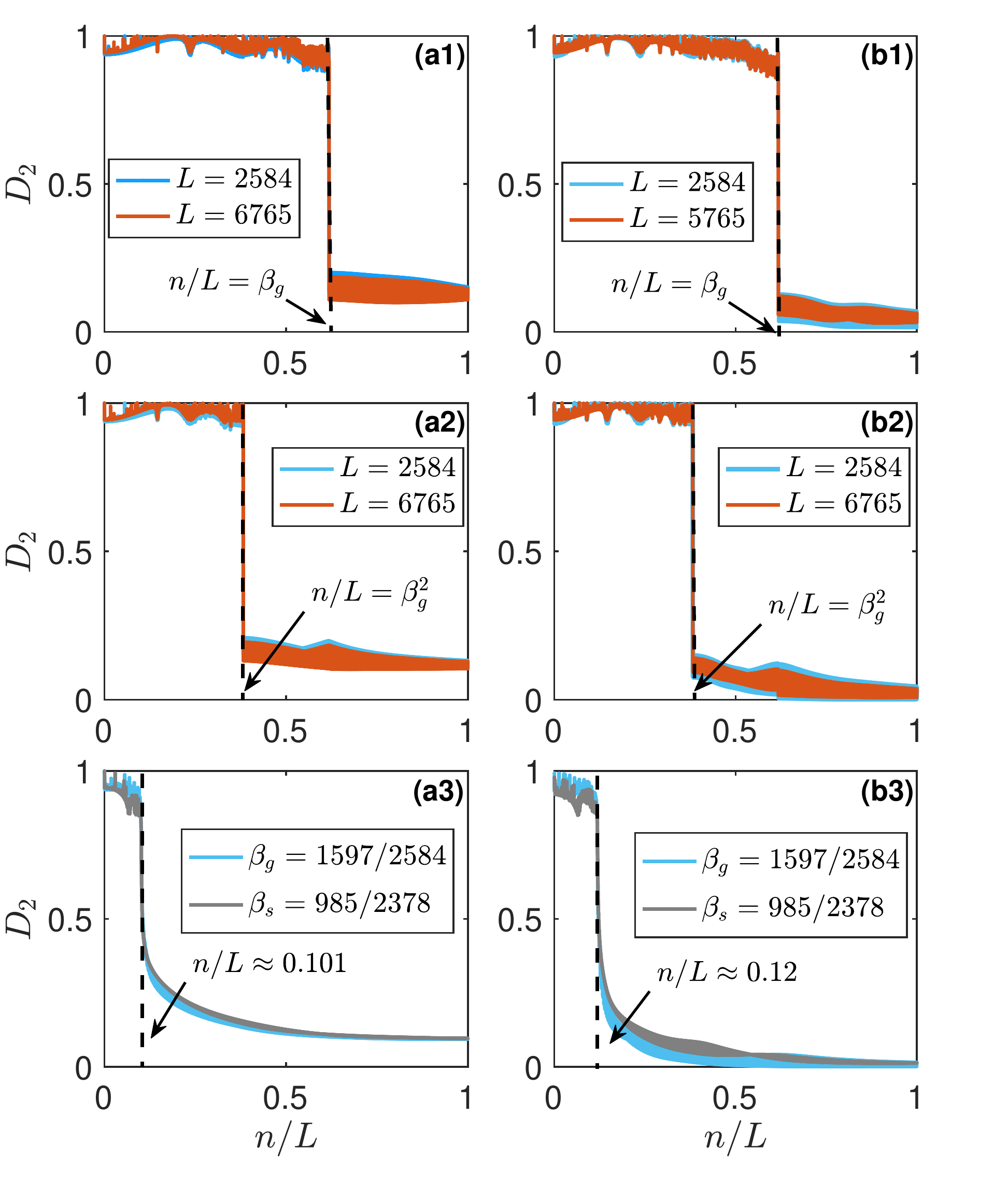}
	
	\renewcommand\figurename{Fig}
	\caption{(a1)-(a3) $D_2$ versus the index $n / L$  for $k=0.8$ and $a=0.5$
	with $\Delta=0.3J$, $0.5J$, and $2J$, respectively. (b1)-(b3) $D_2$ versus the index $n / L$ for $k=0.8$ and $a=2.0$
	with $\Delta$=0.73, 0.9 and 1.6, respectively. The dashed lines represent the energy indexes of the
localization transitions.	Here, for $L=2584$, we take $\beta_g=1597 / 2584$ and $d=4$, and for $L=6765$,
we choose $\beta_g=4181 / 6765$, 	$d=5$, respectively. For $\beta_s=985/2378$, we take $L=2378$ and $d=2$.}\label{f3}
\end{figure}

Next, we further study the different localization phenomena in the long-range hoping and the short-range hopping cases and the differences between the $P_{\ell}$ regimes and the normal intermediate regimes.
We fix $L=2584$ and $\beta_{g}=1597/2584$ in the calculations. For long-range hopping case ($a=0.5$) and $\Delta=0.3$
chosen in the $P_{1}$ regime, we can see that in Fig.~\ref{f3}(a1), below $n/L= \beta_{g}$, $D_{2}$ tends to $1$,
corresponding to the ergodic eigenstates, and above  $n/L= \beta_{g}$, $D_{2}$ tends to a finite value, corresponding
to the multifractal eigenstates. In this case, the abrupt change of $D_{2}$ from $1$ to a non-zero value presents an
EM transition at $n/L=\beta_{g}$. In contrast to
the long-range case, we can see that for short-range hopping (Fig.~\ref{f3}(b1) shows), $D_{2}$ changes from $1$ to zero, showing
an EL transition at $n/L=\beta_{g}$. For higher $P_{\ell}$ regimes, the similar phenomena still exists.
We take $\Delta=0.5$ and $\Delta=0.9$ from the $P_{2}$ regime, the corresponding $D_{2}$ for $a=0.5$ and $a=2$ are plotted
in Fig.~\ref{f3}(a2) and Fig.~\ref{f3}(b2), respectively. The two diagrams present an EM transition and an EL transition
at $n/L=\beta^{2}_{g}$, respectively. As shown in  Figs.~\ref{f3}(a1), (a2), (b1), and (b2), we can see the fractal dimensions $D_2$
are independent on the system size  $L$. From the above analysis, we can see that in the $P_{\ell}$ regimes, the two types of  edges  show dependence on $\beta$.
In fact, the two types of transitions  appear in the normal intermediate regimes as well (see the EM transition in Fig.~\ref{f3}(a3) \
and the EL transition in Fig.~\ref{f3}(b3), respectively), where the EL edge and EM edge are visibly $\beta$ independent. Meanwhile, the results suggest that the features aroused by the hopping types (controlled by $a$) are robust against the weak non-Hermitian effect.

\begin{figure}[htp]
	\begin{minipage}[h]{0.9\linewidth}
	\centering
	\includegraphics[width=1.0\textwidth]{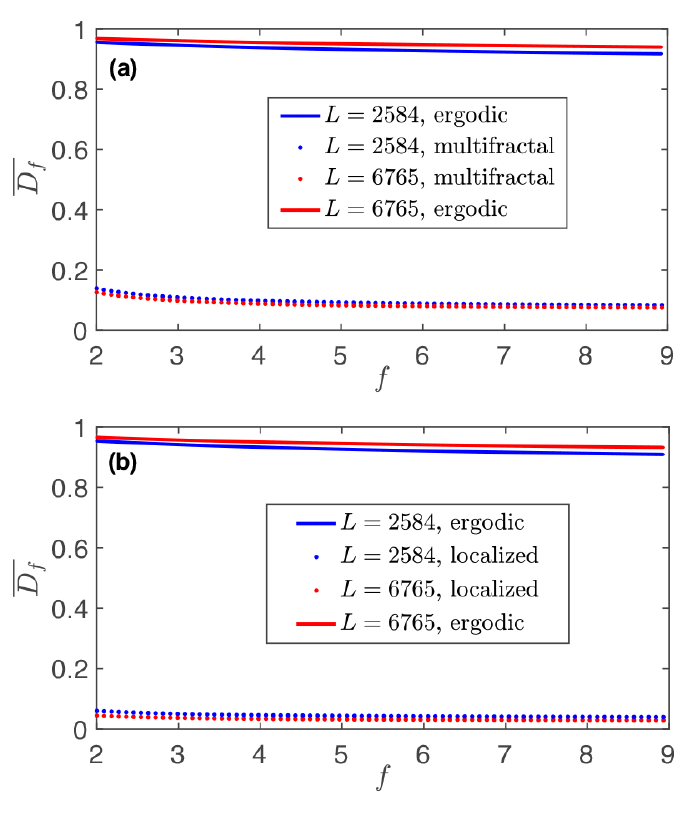}
	\renewcommand\figurename{Fig}
\end{minipage}
	
\renewcommand\figurename{Fig}
\caption{ (a) Averaged fractal dimension $\overline {D_{f}} $ as a function of $f$ for $k=0.8$, $\beta_g =1597/2584$,
$\Delta=0.4J$ and $a= 0.5$ for the system in the $P_{2}$ regime with an EM edge.
(b) Averaged fractal dimension $\overline {D_{f}} $ as a function of $f$ for $k=0.8$, $\beta_g =1597/2584$,
$\Delta=0.9J$ and $a= 2$  for  the system in the $P_{2}$ regime with an EL edge.
$\overline{D_{f}} $ is calculated by averaging over $\beta ^{2}_{g}$	 fraction of ergodic states
and ( $1-\beta ^{2}_{g}$ ) fraction of multifractal/localized states. Here, for $L=2584$, we
take $\beta_g=1597 / 2584$ and $d=4$, and for $L=6765$, we take $\beta_g=4181 / 6765$, $d=5$.}\label{f4}
\end{figure}

In the above analysis, we have used the special fractal dimension $D_2$ to determine the localization properties of the system. To further clarify the existence of mulitifractality in the regime $a < 1$,
we plot the average of $D_{f}$ over the target eigenstates, i.e., $\overline{D_{f}}$, as a function
of $f$ for the $P_2$ regime ($\lambda=\beta _g^2$) for $a=0.5$ [in Fig.~\ref{f4}(a)] and $a=2$ [in Fig.~\ref{f4}(b)]
in the system with $\beta=\beta _g$. In Figs.~\ref{f4}(a) and \ref{f4}(b), $\overline {D_{f}}$ shown with
the solid red and blue curves are the average of $D_{f}$ over the lowest $\beta _g^2L$ eigenstates and are close to 1 for
different $f$, indicating the ergodic states, and are almost independent of the system sizes.
$\overline{D_{f}} $ shown with the blue and red dashed curves are the average of $D_{f}$ over the highest $(1-\beta _g^2)L$ eigenstates. Intuitively, for
various $f$ and system sizes, $\overline{D_{f}}$  show a weak dependence on $f$ for $a = 0.5$, whereas $D_f$ approach
to $0$ and is almost independent of $f$ for $a = 2$.  It indicates that these eigenstates are multifractal for $a = 0.5$ and
localized for $a = 2$.

Now we first discuss the absence of localized states for $a<1$. After performing the Fourier transformation
$g(\tilde{\theta})=\frac{1}{\sqrt{L}}\sum_{j}\phi_{j}e^{i\tilde{\theta}j}$ where $\tilde{\theta}=2\pi\theta$,
we have the following dual equation of Eq.~(\ref{eigen_1})
\begin{equation}\label{eigen_2}
\frac{\Delta e^{k}}{2}g(\tilde{\theta}-\tilde{\omega})+\frac{\Delta e^{-k}}{2}g(\tilde{\theta}+\tilde{\omega})=\left(E+\sum_{s}\frac{2}{s^{a}}\cos(s\tilde{\theta})\right)g(\tilde{\theta}),
\end{equation}
where $\tilde{\omega}=2\pi\beta$ and the index $n$ has been suppressed.
For $a<1$, the dual potential $\sum_{s}2\cos(s\tilde{\theta})/s^{a}$ is divergent. According to Simon-Spencer theorem \cite{simon1989trace,simon2005trace}
and its application \cite{PhysRevB.104.224204}, the spectrum $E$ of the dual eigenfunction is not absolutely continuous. Thus, for our model, there is
no any localized state in the $0<a<1$ regime.

Next, we analyze the location of the intermediate regime and the critical point of
EL transition for $a>1$. Here, we take $a=2$ as a specific example.
According to the Avila's global theory \cite{avila2015global} and its application \cite{cai}, we first
make an analytical continuation on $\Delta_{j}$, i.e., $ik \rightarrow i(k+\delta)$. Thus, in the limit $\delta\rightarrow \infty$,
the dual equation in Eq.~(\ref{eigen_2}) reduces to
\begin{equation}
\frac{\Delta e^{k+\delta}}{2}g(\tilde{\theta}-\tilde{\omega})=\left(E+\sum_{s}\frac{2}{s^{a}}\cos(s\tilde{\theta})\right)g(\tilde{\theta}).
\end{equation}
Ref.~\cite{PhysRevB.104.224204} tells us that we can analytically extract the localization properties when the analytical
continuation $\delta$ recovers to zero. Meanwhile, the infinite series $\sum_{s}2\cos(s\tilde{\theta})/s^{2}$ converges
to ${\tilde{\theta}}^2/2-\pi\tilde{\theta}+\pi^2/3$ (see the derivation in Appendix \ref{A}). Therefore, we finally obtain the
following dual equation
\begin{equation}
\frac{\Delta e^{k}}{2}g(\tilde{\theta}-\tilde{\omega})=\left(E+{\tilde{\theta}}^2/2-\pi\tilde{\theta}+\pi^2/3\right)g(\tilde{\theta}).
\end{equation}

The Sarnak method \cite{sarnak1982spectral} and its application \cite{xu2022exact} tells us that  the location
of the intermediate regime and the EL edge are related to the following characteristic function
\begin{equation}
\begin{aligned}
G(E)&=\frac{1}{2\pi}\int^{2\pi}_{0}\ln\left|E+\frac{\tilde{\theta}^2}{2}-\pi\tilde{\theta}+\frac{\pi^2}{3}\right|\\
&=-\ln2+\frac{1}{2\pi}\int^{2\pi}_{0}\ln\left|\left(\tilde{\theta}-\pi\right)^2-\left(\frac{\pi^2}{3}-2E\right)\right|\\
&=-2-\ln2+\frac{\epsilon_{+}\ln\epsilon_{+}+\epsilon_{-}\ln\epsilon_{-}}{\pi},
\end{aligned}
\end{equation}
where $\epsilon_{\pm}=\pi \pm \sqrt{\pi^2/3-2E}$.
By $\{G(E)>\ln|\frac{\Delta e^{k}}{2}|\}\cap \epsilon_{E}$, where the set of spectrum $\epsilon_{E}=\left[-\pi^2/3,\pi^2/6\right]$
guarantees the existence of the solution to the equation $E+{\tilde{\theta}}^2/2-\pi\tilde{\theta}+\pi^2/3=0$ with $E$ being a real value,
we can locate the intermediate regime. Within $\epsilon_{E}$, we have $G\in\left[2\ln\pi-\ln2-2,2\ln\pi+\ln2-2\right]$. Therefore,
the lower bound of the intermediate regime satisfies $\Delta_{c1}=2e^{-k}e^{2\ln\pi-\ln2-2}$, and the upper bound
satisfies $\Delta_{c2}=2e^{-k}e^{2\ln\pi+\ln2-2}$. When $\Delta<\Delta_{c1}$, all the eigenstates are ergodic, and when
$\Delta>\Delta_{c2}$, all the eigenstates are localized. For the $k=0.8$ case, $\Delta_{c1}\approx 0.6J$
and $\Delta_{c2}\approx 2.4J$, which are shown in Figs.~\ref{f2}(b) and \ref{f2}(d), labeled by blue and white dashed lines, respectively.

\section{The localization properties under the strong non-Hermitian effect}\label{S4}
  In this section, we study the localization properties under the strong non-Hermitian effect with $k=3$.
  We find that the long-range-hopping induced EM edges and the short-range-hopping induced  EL edges are
  robust against the strong non-Hermitian effect, but the $\beta$-dependent $P_{\ell}$ regimes disappear
  completely. Meanwhile, the fractions of the EM and EL edges are completely independent of the
  value of $\beta$.
  \begin{figure}[htp]
		\centering
		\includegraphics[width=9.0cm]{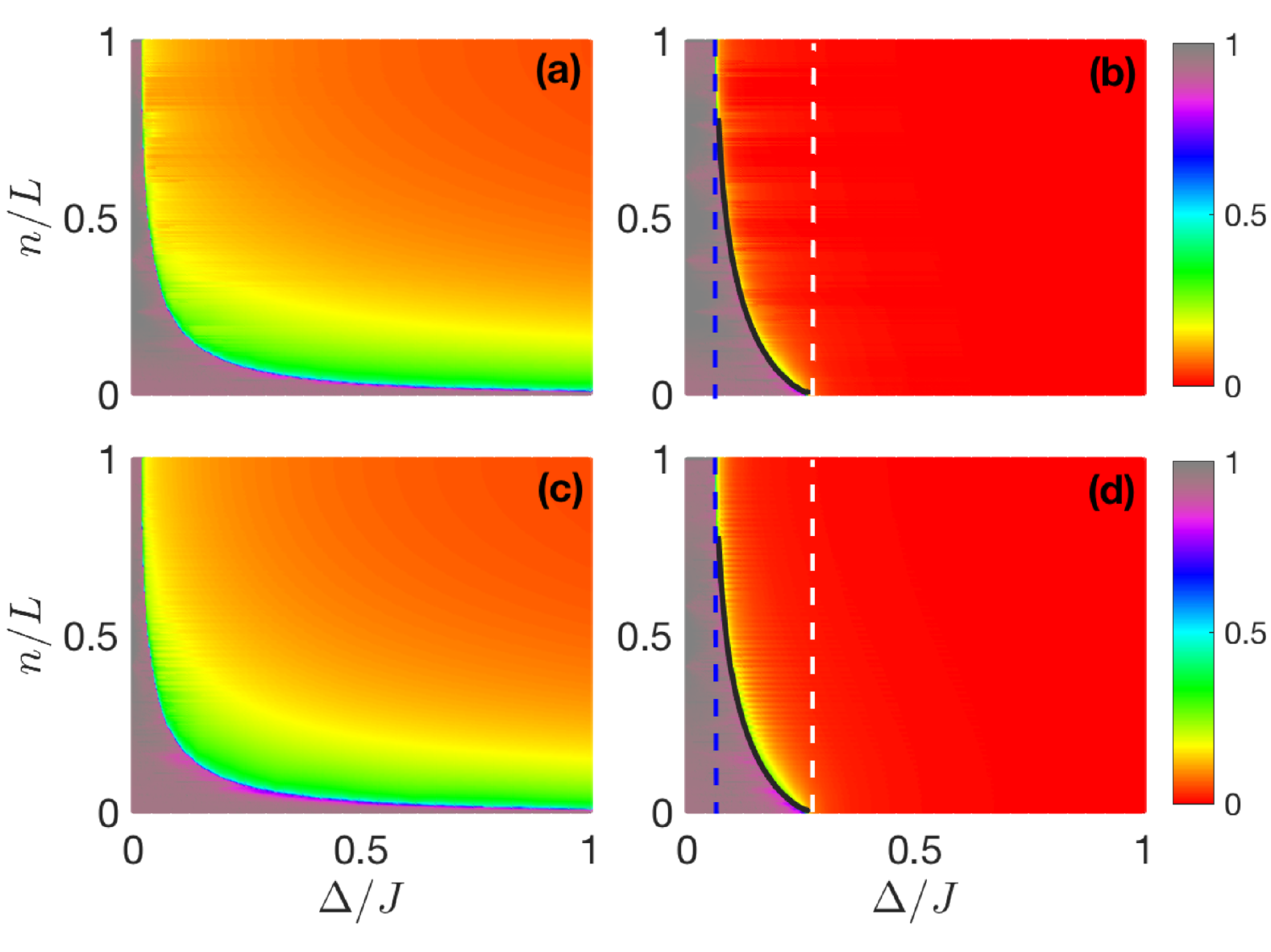}
		\renewcommand\figurename{Fig}
	\renewcommand\figurename{Fig}
	\caption{Fractal dimension $D_2 $ (shown in color) of different eigenstates as a function of the strength of the
	complex potential $\Delta$ with $k=3$, $L=2584$, $d=4$, and $\beta_g =1597/2584$ for (a) $a=0.5$  and (b) $a=2$,
	and with $k=3$, $L=2378$, $d=2$, and $\beta_s =985/2378$ for (c) $a=0.5$ and (d) $a=2$. In (b) and (d), the blue
	and white dashed lines satisfy $\Delta_{c 1} \approx 0.066J$ and $\Delta_{c 2} \approx 0.266J$, respectively, and
	the black solid lines represent the EL edges $E_c$ determined by $G(E_{c})=\ln|\frac{\Delta e^{k}}{2}|$.}\label{f5}
	\end{figure}
	\begin{figure}[htp]
		\centering
		\includegraphics[width=9.5cm]{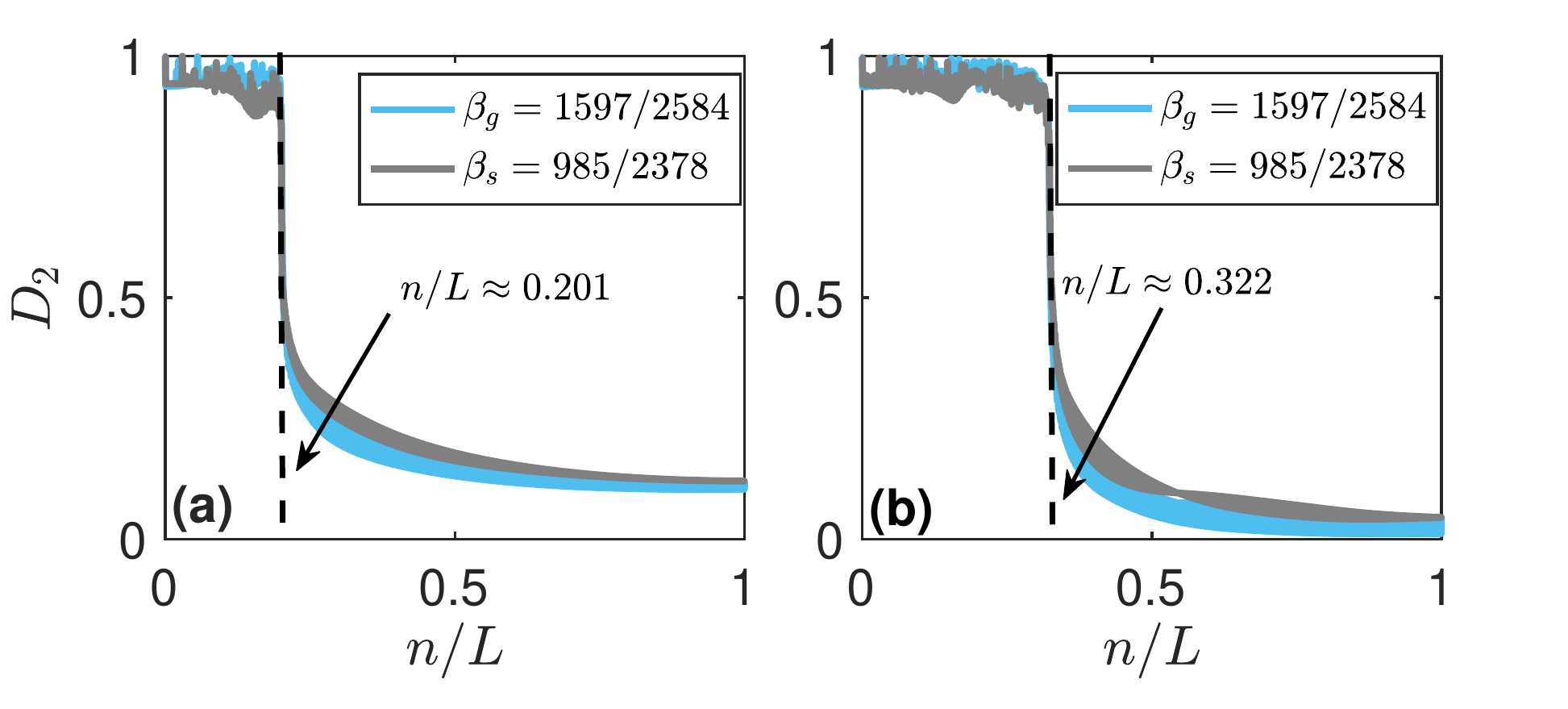}
		\renewcommand\figurename{Fig}
		\caption{
$D_2$ versus the index $n / L$ with different $\beta$  and $k=3$.  (a) $a=0.5$ and $\Delta=0.1J$. (b) $a=2$ and $\Delta=0.11J$.
The dashed lines represent the energy indexes of the localization transitions
Here, for $\beta_g=1597/2584$, we take $L=2584$, $d=4$, and for $\beta_s=985/2378$,
we take $L=2378$ and $d=2$.}\label{f6}
\end{figure}
In order to study the localization properties under the strong non-Hermitian effect with $k=3$,
we calculate the fractal dimension $D_2 $  of different eigenstates as a function of $\Delta$.
As shown in Figs.~\ref{f5}(a) and \ref{f5}(c), for $a=0.5$, no matter $\beta$ is $\beta_g =1597/2584$
or $\beta_s =985/2378$, the EM edge smoothly decreases as $\Delta$ increases, and the EM edges are
$\beta$ independent. As shown in Figs.~\ref{f5}(b) and \ref{f5}(d), for $a=2$, no matter $\beta$ is $\beta_g =1597/2584$
or $\beta_s =985/2378$, the EL edges smoothly decay with the increase of $\Delta$ and is independent of $\beta$, too.	
In addition, the bounds of the intermediate regime for the short-range hopping case ($a=2$) can be analytically obtained as well.
Employing the same analytical methods as those done for the $k=0.8$ case, here the lower bound of the intermediate regime
satisfies $\Delta_{c1}\approx 0.066J$ and the upper bound satisfies $\Delta_{c2}\approx 0.266J$, which are shown in
Figs.~\ref{f5}(b) and \ref{f5}(d), labeled by blue and white dashed lines, respectively. Meanwhile, according to the
above mentioned Sarnak method, the critical point $E_{c}$ of the EL transition can be determined by $G(E_{c})=\ln|\frac{\Delta e^{k}}{2}|$,
which are labeled by the black solid lines in Figs.~\ref{f5}(b) and \ref{f5}(d).
	
To further explain the $\beta$-independent features, we calculate single-parameter $D_{2}$ curves for
two different quasiperiodic  parameter $\beta$. For $a=0.5$, Fig.~\ref{f6}(a)  shows that when $\beta$ are taken
as $\beta_g =1597/2584$ and $\beta_s =985/2378$, the fractal dimension
$D_2 $ both jump at $n/L \approx  0.201$ under the same parameter $a=2$
and $\Delta = 0.1$. When $n/L < 0.201$.	the corresponding eigenstates are ergodic with $D_2 \approx 1$.
For $n/L > 0.201$, the corresponding eigenstates show the multifractal feature with $D_2$ being finite values.
This indicates that there are same EM edges at $n/L \approx  0.201$ for different $\beta$. As shown in Fig. \ref{f6}(b),
no matter $\beta$ is equal to $\beta_g =1597/2584$ or equal to $\beta_s =985/2378$, the fractal dimension $D_2 $
under $k=3$, $\Delta = 0.11J$, and $a=2$ both jump from $D_2 \rightarrow   1$ to $D_2 \rightarrow  0 $ at $n/L \approx  0.322$.
This indicates that there are the same EL edges at $n/L \approx  0.322$  for different $\beta$.


      \begin{figure}[htp]
      \small
	  \begin{minipage}[h]{1.0\linewidth}
		 \centering
		 \includegraphics[width=1.0\textwidth]{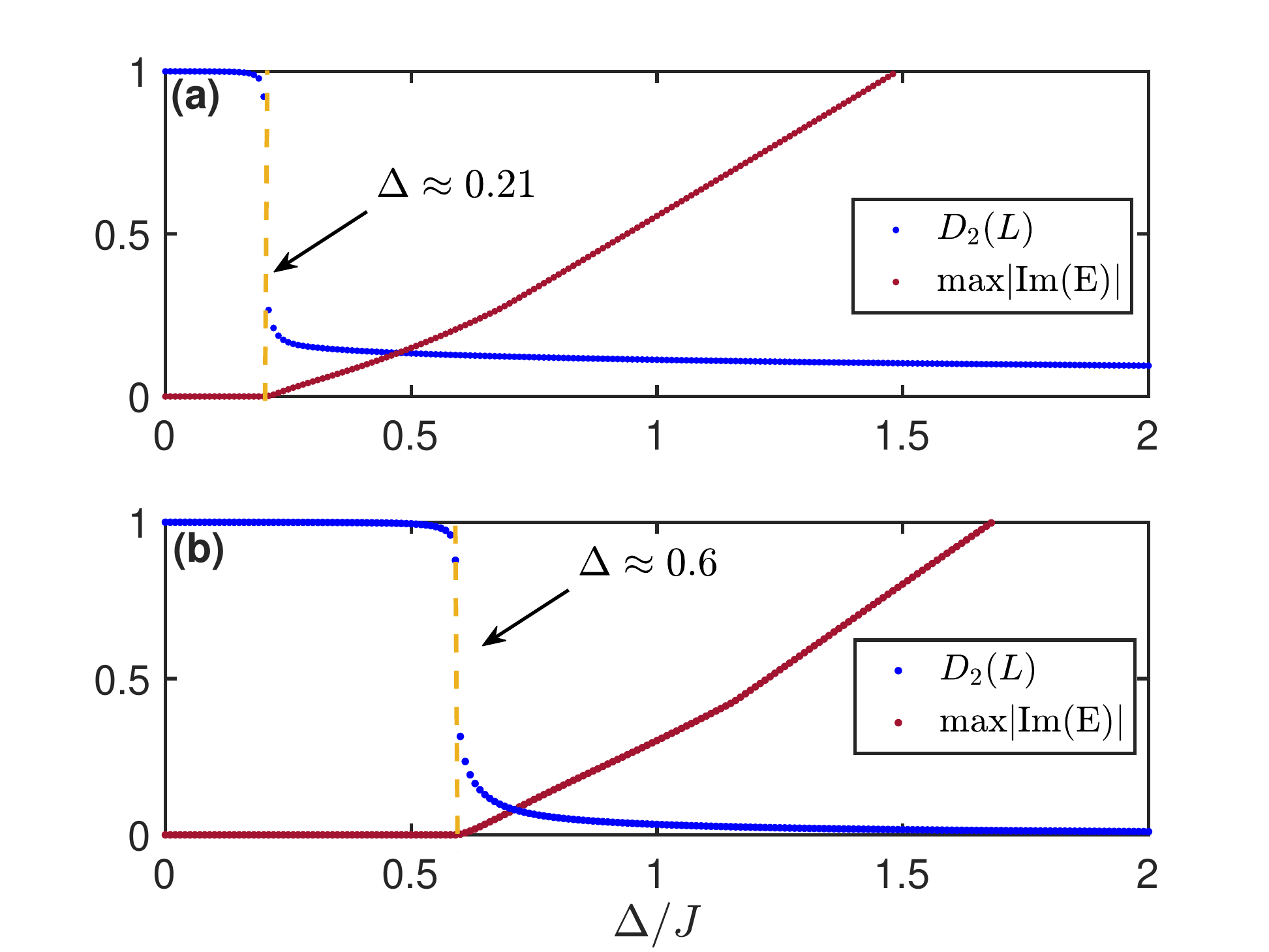}
		 \renewcommand\figurename{Fig}
		\end{minipage}
		\renewcommand\figurename{Fig}
	\caption{The behaviors of the maximum value of $|\rm{Im}(E)|$ and
	the fractal dimension of the $Lth$ eigenstate $D_2{(L)}$ as the functions of $\Delta$ with $k=0.8$,
	$\beta _g = 1597/2584$ and $L=2584$ for (a) $a = 0.5$ and (b) $a=2$ respectively.
	The dashed lines denote the $PT$ symmetry breaking point and the EM transition point in (a) and
	EL transition point in (b), respectively.}\label{f7}
	\end{figure}
	
	 \begin{figure}[htp]
		
		\begin{minipage}[h]{1.0\linewidth}
		\centering
		\includegraphics[width=1.0\textwidth]{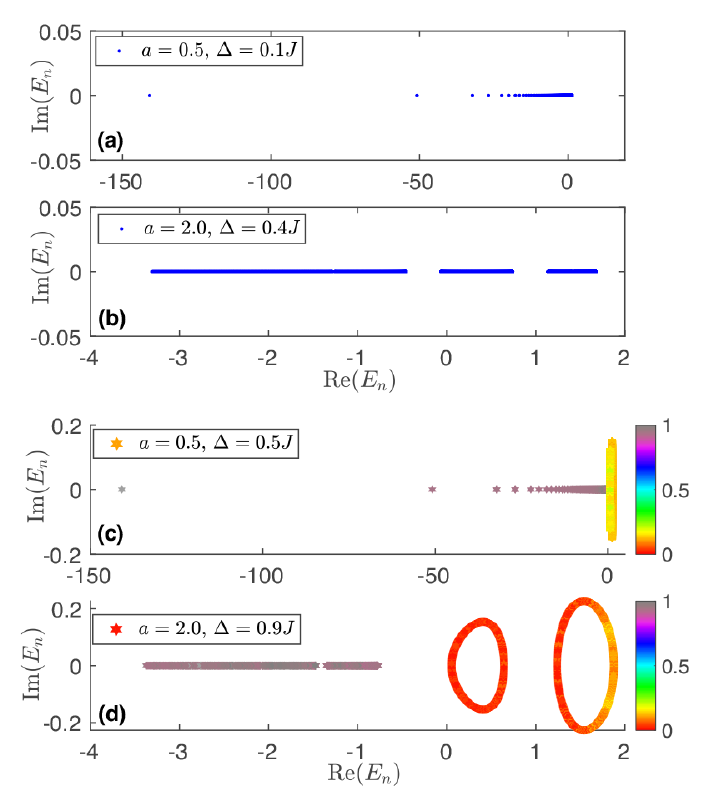}
		\renewcommand\figurename{Fig}
	\end{minipage}
		 \renewcommand\figurename{Fig}
		 \caption{Energy spectrum $E_n$  with $L=2584$,  $k=0.8$, and $\beta _g = 1597/2584 $  for  (a)  $\Delta  = 0.1J$, $a = 0.5$,
	 (b) $\Delta  = 0.4J$, $a = 2$,
		   (c) $a=0.5$, $\Delta  = 0.5J$,  and (d)  $a=2$, $\Delta  = 0.9J$, respectively.
		 The color bar shows the value of fractal dimension $D_2 $. }\label{f8}
	 \end{figure}
	
\section{$PT$ symmetry breaking}\label{S5}
 Next, we study the $PT$ symmetry breaking in the cases of $a<1$  and $a>1$. Figures. \ref{f7}(a) and \ref{f7}(b) present
 the behaviors of the maximum value of $|\operatorname{Im}(E)|$ and the fractal dimension $D_2{(L)}$ of the $L$th eigenstate
 as the function of $\Delta$ for $a=0.5$ and $a=2$, respectively. As can be seen from Fig. \ref{f7}(a), the $PT$ symmetry breaking
 point coincides with the EM phase transition point at $\Delta \approx 0.21J$ for $a=0.5$, and coincides with the EL phase transition
 point at $\Delta \approx 0.6J$ for $a=2$ [in Fig. \ref{f7}(b)]. The energy spectrum for $\Delta=0.1J$ and $\Delta=0.4J$ are shown
 in Fig. \ref{f8}(a) and \ref{f8}(b), respectively, where all the eigenvalues are real. In the intermediate regime, the complex energies emerge.
 As shown in Fig. \ref{f8}(c) with $a=0.5$ and  $\Delta=0.5J$, the real-complex transition of the energy spectrum is synchronized
 with the EM transition [see Fig.~\ref{f3}(a2)]. Besides, we can see that in Fig. \ref{f8}(d) with $a=2$ and  $\Delta=0.9J$, there
 exist the real-complex transition of the energy spectrum accompanied by the EL transition [see Fig.~\ref{f3}(b2)].
\section{ conclusion}\label{S6}
In conclusion, a non-Hermitian AAH model with power-law hopppings was studied. We uncover that the quasiperidic parameter
$\beta$-dependent $P_{\ell}$ regimes are robust  against the weak non-Hermitian effect. When the non-Hermtian effect
gets stronger, the $P_{\ell}$ regimes disappear. However, we find that localization properties, i.e., the long-range hopping induced
EM edge and the short-range hopping induced EL edge are robust against the non-Hermitian effect, and
are well characterized by the fractal dimension $D_{2}$. We argued that the
absence of the localized states for the long-range hopping case by the Simon-Spencer theorem. Meanwhile, by employing the Sarnak method and
Avila's global theory, the boundaries of the intermediate regime and the critical points of the EL  phase transition for the short-range hopping case ($a=2$)
are analytically located, which are coincident with the numerical results. Finally, we analyzed the relationship between the PT symmetry breaking
and the EM and EL phase transitions. We found that the ergodic eigenstates correspond to the real eigenenergies, whereas the multifractal and
localized eigenstates correspond to the complex eigenenergies.

The authors acknowledge support from NSFC under
Grants No. 11835011 and No. 12174346.
\appendix
\section{The derivation of $\sum_{s}2\cos(s\tilde{\theta})/s^{2}={\tilde{\theta}}^2/2-\pi\tilde{\theta}+\pi^2/3$ }\label{A}
According to the equation:
\begin{equation}
	\begin{aligned}
\ln\left(1-x\right)=-\sum_{n = 1}^{\infty} \frac{x^n}{n},
\end{aligned}
\end{equation}
 we have
\begin{equation}
\begin{aligned}
 &\sum_{n = 1}^{\infty} \frac{e^{inx}}{n}=-\ln\left(1-e^{ix}\right)\\
 &=\sum_{n = 1}^{\infty} \frac{\cos\left(nx\right)}{n}+i\sum_{n = 1}^{\infty} \frac{\sin\left(nx\right)}{n},
\end{aligned}
\end{equation}
 where
 \begin{equation}
	\begin{aligned}
 \ln\left(1-e^{ix}\right)=\ln||1-e^{ix}||+i{\rm Arg}\left(1-e^{ix}\right)
\end{aligned}
\end{equation}
 and
\begin{equation}\label{A4}
\begin{aligned}
 &\sum_{n = 1}^{\infty} \frac{\sin\left(nx\right)}{n}=-{\rm Im}\left[\ln\left(1-e^{ix}\right)\right] \\
 &=-{\rm Arg}\left(1-e^{ix}\right)=-\arctan\left\{\frac{-\sin\left(x\right)}{1-\cos\left(x\right)}\right\}\\
 &=-\arctan\left\{\frac{\left[-2\tan{\left(\frac{x}{2}\right)}\right]/\left[1+\tan^{2}{\left(\frac{x}{2}\right)}\right]}{1-\left[1-\tan^{2}{\left(\frac{x}{2}\right)}\right]/\left[1+\tan^{2}{\left(\frac{x}{2}\right)}\right]}\right\}\\
&=\arctan\left\{\frac{1}{\tan\left(\frac{x}{2}\right) } \right\}=\frac{\pi}{2} -\frac{x}{2}.
\end{aligned}
\end{equation}
Replacing $n$ with $s$, and $x$ with $\tilde{\theta}$, Eq.~(\ref{A4}) reads
\begin{equation}\label{A5}
\sum_{s = 1}^{\infty} \frac{\sin(s\tilde{\theta})}{s}=\frac{\pi}{2} -\frac{\tilde{\theta}}{2}.
\end{equation}
After making an integration on $\tilde{\theta}$ in Eq.~(\ref{A5}), we have
\begin{equation}
 -\sum_{s = 1}^{\infty} \frac{\cos(s\tilde{\theta})}{s^2}=\frac{\pi\tilde{\theta}}{2} -\frac{\tilde{\theta}^2}{4}+{\rm const}.
\end{equation}
When $\tilde{\theta}=0$, we get
 \begin{equation}
	\begin{aligned}
  {\rm const}=-\sum_{s = 1}^{\infty} \frac{1}{s^2}=-\frac{\pi^2}{6}
\end{aligned}
\end{equation}
 Finally, we obtain
  \begin{equation}
	\begin{aligned}
\sum_{s = 1}^{\infty} \frac{2\cos(s\tilde{\theta})}{s^2}=\frac{\tilde{\theta}^2}{2}-{\pi\tilde{\theta}} +\frac{\pi^2}{3}.
\end{aligned}
\end{equation}
\newpage
	\bibliography{re}

\begin{thebibliography}{52}%
\makeatletter
\providecommand \@ifxundefined [1]{%
 \@ifx{#1\undefined}
}%
\providecommand \@ifnum [1]{%
 \ifnum #1\expandafter \@firstoftwo
 \else \expandafter \@secondoftwo
 \fi
}%
\providecommand \@ifx [1]{%
 \ifx #1\expandafter \@firstoftwo
 \else \expandafter \@secondoftwo
 \fi
}%
\providecommand \natexlab [1]{#1}%
\providecommand \enquote  [1]{``#1''}%
\providecommand \bibnamefont  [1]{#1}%
\providecommand \bibfnamefont [1]{#1}%
\providecommand \citenamefont [1]{#1}%
\providecommand \href@noop [0]{\@secondoftwo}%
\providecommand \href [0]{\begingroup \@sanitize@url \@href}%
\providecommand \@href[1]{\@@startlink{#1}\@@href}%
\providecommand \@@href[1]{\endgroup#1\@@endlink}%
\providecommand \@sanitize@url [0]{\catcode `\\12\catcode `\$12\catcode
  `\&12\catcode `\#12\catcode `\^12\catcode `\_12\catcode `\%12\relax}%
\providecommand \@@startlink[1]{}%
\providecommand \@@endlink[0]{}%
\providecommand \url  [0]{\begingroup\@sanitize@url \@url }%
\providecommand \@url [1]{\endgroup\@href {#1}{\urlprefix }}%
\providecommand \urlprefix  [0]{URL }%
\providecommand \Eprint [0]{\href }%
\providecommand \doibase [0]{http://dx.doi.org/}%
\providecommand \selectlanguage [0]{\@gobble}%
\providecommand \bibinfo  [0]{\@secondoftwo}%
\providecommand \bibfield  [0]{\@secondoftwo}%
\providecommand \translation [1]{[#1]}%
\providecommand \BibitemOpen [0]{}%
\providecommand \bibitemStop [0]{}%
\providecommand \bibitemNoStop [0]{.\EOS\space}%
\providecommand \EOS [0]{\spacefactor3000\relax}%
\providecommand \BibitemShut  [1]{\csname bibitem#1\endcsname}%
\let\auto@bib@innerbib\@empty
\bibitem [{\citenamefont {Anderson}(1958)}]{PhysRev.109.1492}%
  \BibitemOpen
  \bibfield  {author} {\bibinfo {author} {\bibfnamefont {P.~W.}\ \bibnamefont
  {Anderson}},\ }\bibfield  {title} {\enquote {\bibinfo {title} {Absence of
  diffusion in certain random lattices},}\ }\href {\doibase
  10.1103/PhysRev.109.1492} {\bibfield  {journal} {\bibinfo  {journal} {Phys.
  Rev.}\ }\textbf {\bibinfo {volume} {109}},\ \bibinfo {pages} {1492} (\bibinfo
  {year} {1958})}\BibitemShut {NoStop}%
\bibitem [{\citenamefont {Abrahams}\ \emph {et~al.}(1979)\citenamefont
  {Abrahams}, \citenamefont {Anderson}, \citenamefont {Licciardello},\ and\
  \citenamefont {Ramakrishnan}}]{PhysRevLett.42.673}%
  \BibitemOpen
  \bibfield  {author} {\bibinfo {author} {\bibfnamefont {E.}~\bibnamefont
  {Abrahams}}, \bibinfo {author} {\bibfnamefont {P.~W.}\ \bibnamefont
  {Anderson}}, \bibinfo {author} {\bibfnamefont {D.~C.}\ \bibnamefont
  {Licciardello}}, \ and\ \bibinfo {author} {\bibfnamefont {T.~V.}\
  \bibnamefont {Ramakrishnan}},\ }\bibfield  {title} {\enquote {\bibinfo
  {title} {Scaling theory of localization: Absence of quantum diffusion in two
  dimensions},}\ }\href {\doibase 10.1103/PhysRevLett.42.673} {\bibfield
  {journal} {\bibinfo  {journal} {Phys. Rev. Lett.}\ }\textbf {\bibinfo
  {volume} {42}},\ \bibinfo {pages} {673} (\bibinfo {year} {1979})}\BibitemShut
  {NoStop}%
\bibitem [{\citenamefont {Lahini}\ \emph {et~al.}(2008)\citenamefont {Lahini},
  \citenamefont {Avidan}, \citenamefont {Pozzi}, \citenamefont {Sorel},
  \citenamefont {Morandotti}, \citenamefont {Christodoulides},\ and\
  \citenamefont {Silberberg}}]{PhysRevLett.100.013906}%
  \BibitemOpen
  \bibfield  {author} {\bibinfo {author} {\bibfnamefont {Y.}~\bibnamefont
  {Lahini}}, \bibinfo {author} {\bibfnamefont {A.}~\bibnamefont {Avidan}},
  \bibinfo {author} {\bibfnamefont {F.}~\bibnamefont {Pozzi}}, \bibinfo
  {author} {\bibfnamefont {M.}~\bibnamefont {Sorel}}, \bibinfo {author}
  {\bibfnamefont {R.}~\bibnamefont {Morandotti}}, \bibinfo {author}
  {\bibfnamefont {D.~N.}\ \bibnamefont {Christodoulides}}, \ and\ \bibinfo
  {author} {\bibfnamefont {Y.}~\bibnamefont {Silberberg}},\ }\bibfield  {title}
  {\enquote {\bibinfo {title} {Anderson localization and nonlinearity in
  one-dimensional disordered photonic lattices},}\ }\href {\doibase
  10.1103/PhysRevLett.100.013906} {\bibfield  {journal} {\bibinfo  {journal}
  {Phys. Rev. Lett.}\ }\textbf {\bibinfo {volume} {100}},\ \bibinfo {pages}
  {013906} (\bibinfo {year} {2008})}\BibitemShut {NoStop}%
\bibitem [{\citenamefont {Lahini}\ \emph {et~al.}(2010)\citenamefont {Lahini},
  \citenamefont {Bromberg}, \citenamefont {Christodoulides},\ and\
  \citenamefont {Silberberg}}]{PhysRevLett.105.163905}%
  \BibitemOpen
  \bibfield  {author} {\bibinfo {author} {\bibfnamefont {Y.}~\bibnamefont
  {Lahini}}, \bibinfo {author} {\bibfnamefont {Y.}~\bibnamefont {Bromberg}},
  \bibinfo {author} {\bibfnamefont {D.~N.}\ \bibnamefont {Christodoulides}}, \
  and\ \bibinfo {author} {\bibfnamefont {Y.}~\bibnamefont {Silberberg}},\
  }\bibfield  {title} {\enquote {\bibinfo {title} {Quantum correlations in
  two-particle anderson localization},}\ }\href {\doibase
  10.1103/PhysRevLett.105.163905} {\bibfield  {journal} {\bibinfo  {journal}
  {Phys. Rev. Lett.}\ }\textbf {\bibinfo {volume} {105}},\ \bibinfo {pages}
  {163905} (\bibinfo {year} {2010})}\BibitemShut {NoStop}%
\bibitem [{\citenamefont {Aubry}\ and\ \citenamefont
  {Andr{\'e}}(1980)}]{aubry1980analyticity}%
  \BibitemOpen
  \bibfield  {author} {\bibinfo {author} {\bibfnamefont {S.}~\bibnamefont
  {Aubry}}\ and\ \bibinfo {author} {\bibfnamefont {G.}~\bibnamefont
  {Andr{\'e}}},\ }\bibfield  {title} {\enquote {\bibinfo {title} {Analyticity
  breaking and anderson localization in incommensurate lattices},}\ }\href@noop
  {} {\bibfield  {journal} {\bibinfo  {journal} {Ann. Israel Phys. Soc}\
  }\textbf {\bibinfo {volume} {3}},\ \bibinfo {pages} {18} (\bibinfo {year}
  {1980})}\BibitemShut {NoStop}%
\bibitem [{\citenamefont {{Suslov}}(1982)}]{1982JETP...56..612S}%
  \BibitemOpen
  \bibfield  {author} {\bibinfo {author} {\bibfnamefont {I.~M.}\ \bibnamefont
  {{Suslov}}},\ }\bibfield  {title} {\enquote {\bibinfo {title} {Localization
  in one-dimensional incommensurate systems},}\ }\href
  {https://ui.adsabs.harvard.edu/abs/1982JETP...56..612S} {\bibfield  {journal}
  {\bibinfo  {journal} {Soviet Journal of Experimental and Theoretical
  Physics}\ }\textbf {\bibinfo {volume} {56}},\ \bibinfo {pages} {612}
  (\bibinfo {year} {1982})}\BibitemShut {NoStop}%
\bibitem [{\citenamefont {Wilkinson}(1984)}]{wilkinson1984critical}%
  \BibitemOpen
  \bibfield  {author} {\bibinfo {author} {\bibfnamefont {M.}~\bibnamefont
  {Wilkinson}},\ }\bibfield  {title} {\enquote {\bibinfo {title} {Critical
  properties of electron eigenstates in incommensurate systems},}\ }\href
  {https://doi.org/10.1098/rspa.1984.0016} {\bibfield  {journal} {\bibinfo
  {journal} {Proc. R. Soc. Lond. A}\ }\textbf {\bibinfo {volume} {391}},\
  \bibinfo {pages} {305--350} (\bibinfo {year} {1984})}\BibitemShut {NoStop}%
\bibitem [{\citenamefont {Biddle}\ and\ \citenamefont
  {Das~Sarma}(2010)}]{PhysRevLett.104.070601}%
  \BibitemOpen
  \bibfield  {author} {\bibinfo {author} {\bibfnamefont {J.}~\bibnamefont
  {Biddle}}\ and\ \bibinfo {author} {\bibfnamefont {S.}~\bibnamefont
  {Das~Sarma}},\ }\bibfield  {title} {\enquote {\bibinfo {title} {Predicted
  mobility edges in one-dimensional incommensurate optical lattices: An exactly
  solvable model of anderson localization},}\ }\href {\doibase
  10.1103/PhysRevLett.104.070601} {\bibfield  {journal} {\bibinfo  {journal}
  {Phys. Rev. Lett.}\ }\textbf {\bibinfo {volume} {104}},\ \bibinfo {pages}
  {070601} (\bibinfo {year} {2010})}\BibitemShut {NoStop}%
\bibitem [{\citenamefont {Chen}\ \emph {et~al.}(2022)\citenamefont {Chen},
  \citenamefont {Cheng}, \citenamefont {Lin}, \citenamefont {Asgari},\ and\
  \citenamefont {Xianlong}}]{PhysRevB.106.144208}%
  \BibitemOpen
  \bibfield  {author} {\bibinfo {author} {\bibfnamefont {W.}~\bibnamefont
  {Chen}}, \bibinfo {author} {\bibfnamefont {S.}~\bibnamefont {Cheng}},
  \bibinfo {author} {\bibfnamefont {J.}~\bibnamefont {Lin}}, \bibinfo {author}
  {\bibfnamefont {R.}~\bibnamefont {Asgari}}, \ and\ \bibinfo {author}
  {\bibfnamefont {G.}~\bibnamefont {Xianlong}},\ }\bibfield  {title} {\enquote
  {\bibinfo {title} {Breakdown of the correspondence between the real-complex
  and delocalization-localization transitions in non-hermitian
  quasicrystals},}\ }\href {\doibase 10.1103/PhysRevB.106.144208} {\bibfield
  {journal} {\bibinfo  {journal} {Phys. Rev. B}\ }\textbf {\bibinfo {volume}
  {106}},\ \bibinfo {pages} {144208} (\bibinfo {year} {2022})}\BibitemShut
  {NoStop}%
\bibitem [{\citenamefont {Deng}\ \emph {et~al.}(2019)\citenamefont {Deng},
  \citenamefont {Ray}, \citenamefont {Sinha}, \citenamefont {Shlyapnikov},\
  and\ \citenamefont {Santos}}]{PhysRevLett.123.025301}%
  \BibitemOpen
  \bibfield  {author} {\bibinfo {author} {\bibfnamefont {X.}~\bibnamefont
  {Deng}}, \bibinfo {author} {\bibfnamefont {S.}~\bibnamefont {Ray}}, \bibinfo
  {author} {\bibfnamefont {S.}~\bibnamefont {Sinha}}, \bibinfo {author}
  {\bibfnamefont {G.~V.}\ \bibnamefont {Shlyapnikov}}, \ and\ \bibinfo {author}
  {\bibfnamefont {L.}~\bibnamefont {Santos}},\ }\bibfield  {title} {\enquote
  {\bibinfo {title} {One-dimensional quasicrystals with power-law hopping},}\
  }\href {\doibase 10.1103/PhysRevLett.123.025301} {\bibfield  {journal}
  {\bibinfo  {journal} {Phys. Rev. Lett.}\ }\textbf {\bibinfo {volume} {123}},\
  \bibinfo {pages} {025301} (\bibinfo {year} {2019})}\BibitemShut {NoStop}%
\bibitem [{\citenamefont {Roy}\ and\ \citenamefont
  {Sharma}(2021)}]{PhysRevB.103.075124}%
  \BibitemOpen
  \bibfield  {author} {\bibinfo {author} {\bibfnamefont {N.}~\bibnamefont
  {Roy}}\ and\ \bibinfo {author} {\bibfnamefont {A.}~\bibnamefont {Sharma}},\
  }\bibfield  {title} {\enquote {\bibinfo {title} {Fraction of delocalized
  eigenstates in the long-range aubry-andr\'e-harper model},}\ }\href {\doibase
  10.1103/PhysRevB.103.075124} {\bibfield  {journal} {\bibinfo  {journal}
  {Phys. Rev. B}\ }\textbf {\bibinfo {volume} {103}},\ \bibinfo {pages}
  {075124} (\bibinfo {year} {2021})}\BibitemShut {NoStop}%
\bibitem [{\citenamefont {Liu}\ \emph {et~al.}(2020)\citenamefont {Liu},
  \citenamefont {Guo}, \citenamefont {Pu},\ and\ \citenamefont
  {Longhi}}]{PhysRevB.102.024205}%
  \BibitemOpen
  \bibfield  {author} {\bibinfo {author} {\bibfnamefont {T.}~\bibnamefont
  {Liu}}, \bibinfo {author} {\bibfnamefont {H.}~\bibnamefont {Guo}}, \bibinfo
  {author} {\bibfnamefont {Y.}~\bibnamefont {Pu}}, \ and\ \bibinfo {author}
  {\bibfnamefont {S.}~\bibnamefont {Longhi}},\ }\bibfield  {title} {\enquote
  {\bibinfo {title} {Generalized aubry-andr\'e self-duality and mobility edges
  in non-hermitian quasiperiodic lattices},}\ }\href {\doibase
  10.1103/PhysRevB.102.024205} {\bibfield  {journal} {\bibinfo  {journal}
  {Phys. Rev. B}\ }\textbf {\bibinfo {volume} {102}},\ \bibinfo {pages}
  {024205} (\bibinfo {year} {2020})}\BibitemShut {NoStop}%
\bibitem [{\citenamefont {Biddle}\ \emph {et~al.}(2011)\citenamefont {Biddle},
  \citenamefont {Priour}, \citenamefont {Wang},\ and\ \citenamefont
  {Das~Sarma}}]{PhysRevB.83.075105}%
  \BibitemOpen
  \bibfield  {author} {\bibinfo {author} {\bibfnamefont {J.}~\bibnamefont
  {Biddle}}, \bibinfo {author} {\bibfnamefont {D.~J.}\ \bibnamefont {Priour}},
  \bibinfo {author} {\bibfnamefont {B.}~\bibnamefont {Wang}}, \ and\ \bibinfo
  {author} {\bibfnamefont {S.}~\bibnamefont {Das~Sarma}},\ }\bibfield  {title}
  {\enquote {\bibinfo {title} {Localization in one-dimensional lattices with
  non-nearest-neighbor hopping: Generalized anderson and aubry-andr\'e
  models},}\ }\href {\doibase 10.1103/PhysRevB.83.075105} {\bibfield  {journal}
  {\bibinfo  {journal} {Phys. Rev. B}\ }\textbf {\bibinfo {volume} {83}},\
  \bibinfo {pages} {075105} (\bibinfo {year} {2011})}\BibitemShut {NoStop}%
\bibitem [{\citenamefont {Liu}\ \emph {et~al.}(2021{\natexlab{a}})\citenamefont
  {Liu}, \citenamefont {Wang}, \citenamefont {Zheng},\ and\ \citenamefont
  {Chen}}]{PhysRevB.103.134208}%
  \BibitemOpen
  \bibfield  {author} {\bibinfo {author} {\bibfnamefont {Y.}~\bibnamefont
  {Liu}}, \bibinfo {author} {\bibfnamefont {Y.}~\bibnamefont {Wang}}, \bibinfo
  {author} {\bibfnamefont {Z.}~\bibnamefont {Zheng}}, \ and\ \bibinfo {author}
  {\bibfnamefont {S.}~\bibnamefont {Chen}},\ }\bibfield  {title} {\enquote
  {\bibinfo {title} {Exact non-hermitian mobility edges in one-dimensional
  quasicrystal lattice with exponentially decaying hopping and its dual
  lattice},}\ }\href {\doibase 10.1103/PhysRevB.103.134208} {\bibfield
  {journal} {\bibinfo  {journal} {Phys. Rev. B}\ }\textbf {\bibinfo {volume}
  {103}},\ \bibinfo {pages} {134208} (\bibinfo {year}
  {2021}{\natexlab{a}})}\BibitemShut {NoStop}%
\bibitem [{\citenamefont {Guo}\ \emph {et~al.}(2021)\citenamefont {Guo},
  \citenamefont {Liu}, \citenamefont {Zhao}, \citenamefont {Liu},\ and\
  \citenamefont {Chen}}]{PhysRevLett.127.116801}%
  \BibitemOpen
  \bibfield  {author} {\bibinfo {author} {\bibfnamefont {C.-X.}\ \bibnamefont
  {Guo}}, \bibinfo {author} {\bibfnamefont {C.-H.}\ \bibnamefont {Liu}},
  \bibinfo {author} {\bibfnamefont {X.-M.}\ \bibnamefont {Zhao}}, \bibinfo
  {author} {\bibfnamefont {Y.}~\bibnamefont {Liu}}, \ and\ \bibinfo {author}
  {\bibfnamefont {S.}~\bibnamefont {Chen}},\ }\bibfield  {title} {\enquote
  {\bibinfo {title} {Exact solution of non-hermitian systems with generalized
  boundary conditions: Size-dependent boundary effect and fragility of the skin
  effect},}\ }\href {\doibase 10.1103/PhysRevLett.127.116801} {\bibfield
  {journal} {\bibinfo  {journal} {Phys. Rev. Lett.}\ }\textbf {\bibinfo
  {volume} {127}},\ \bibinfo {pages} {116801} (\bibinfo {year}
  {2021})}\BibitemShut {NoStop}%
\bibitem [{\citenamefont {Liu}\ \emph {et~al.}(2021{\natexlab{b}})\citenamefont
  {Liu}, \citenamefont {Zeng}, \citenamefont {Li},\ and\ \citenamefont
  {Chen}}]{PhysRevB.104.085401}%
  \BibitemOpen
  \bibfield  {author} {\bibinfo {author} {\bibfnamefont {Y.}~\bibnamefont
  {Liu}}, \bibinfo {author} {\bibfnamefont {Y.}~\bibnamefont {Zeng}}, \bibinfo
  {author} {\bibfnamefont {L.}~\bibnamefont {Li}}, \ and\ \bibinfo {author}
  {\bibfnamefont {S.}~\bibnamefont {Chen}},\ }\bibfield  {title} {\enquote
  {\bibinfo {title} {Exact solution of the single impurity problem in
  nonreciprocal lattices: Impurity-induced size-dependent non-hermitian skin
  effect},}\ }\href {\doibase 10.1103/PhysRevB.104.085401} {\bibfield
  {journal} {\bibinfo  {journal} {Phys. Rev. B}\ }\textbf {\bibinfo {volume}
  {104}},\ \bibinfo {pages} {085401} (\bibinfo {year}
  {2021}{\natexlab{b}})}\BibitemShut {NoStop}%
\bibitem [{\citenamefont {Liu}\ \emph {et~al.}(2021{\natexlab{c}})\citenamefont
  {Liu}, \citenamefont {Wang}, \citenamefont {Liu}, \citenamefont {Zhou},\ and\
  \citenamefont {Chen}}]{PhysRevB.103.014203}%
  \BibitemOpen
  \bibfield  {author} {\bibinfo {author} {\bibfnamefont {Y.}~\bibnamefont
  {Liu}}, \bibinfo {author} {\bibfnamefont {Y.}~\bibnamefont {Wang}}, \bibinfo
  {author} {\bibfnamefont {X.-J.}\ \bibnamefont {Liu}}, \bibinfo {author}
  {\bibfnamefont {Q.}~\bibnamefont {Zhou}}, \ and\ \bibinfo {author}
  {\bibfnamefont {S.}~\bibnamefont {Chen}},\ }\bibfield  {title} {\enquote
  {\bibinfo {title} {Exact mobility edges, $\mathcal{PT}$-symmetry breaking,
  and skin effect in one-dimensional non-hermitian quasicrystals},}\ }\href
  {\doibase 10.1103/PhysRevB.103.014203} {\bibfield  {journal} {\bibinfo
  {journal} {Phys. Rev. B}\ }\textbf {\bibinfo {volume} {103}},\ \bibinfo
  {pages} {014203} (\bibinfo {year} {2021}{\natexlab{c}})}\BibitemShut
  {NoStop}%
\bibitem [{\citenamefont {Saha}\ \emph {et~al.}(2019)\citenamefont {Saha},
  \citenamefont {Maiti},\ and\ \citenamefont
  {Purkayastha}}]{PhysRevB.100.174201}%
  \BibitemOpen
  \bibfield  {author} {\bibinfo {author} {\bibfnamefont {M.}~\bibnamefont
  {Saha}}, \bibinfo {author} {\bibfnamefont {S.~K.}\ \bibnamefont {Maiti}}, \
  and\ \bibinfo {author} {\bibfnamefont {A.}~\bibnamefont {Purkayastha}},\
  }\bibfield  {title} {\enquote {\bibinfo {title} {Anomalous transport through
  algebraically localized states in one dimension},}\ }\href {\doibase
  10.1103/PhysRevB.100.174201} {\bibfield  {journal} {\bibinfo  {journal}
  {Phys. Rev. B}\ }\textbf {\bibinfo {volume} {100}},\ \bibinfo {pages}
  {174201} (\bibinfo {year} {2019})}\BibitemShut {NoStop}%
\bibitem [{\citenamefont {Cheng}\ and\ \citenamefont
  {Xianlong}(2022)}]{Cheng_2022}%
  \BibitemOpen
  \bibfield  {author} {\bibinfo {author} {\bibfnamefont {S.}~\bibnamefont
  {Cheng}}\ and\ \bibinfo {author} {\bibfnamefont {G.}~\bibnamefont
  {Xianlong}},\ }\bibfield  {title} {\enquote {\bibinfo {title} {Majorana zero
  modes, unconventional real–complex transition, and mobility edges in a
  one-dimensional non-hermitian quasi-periodic lattice},}\ }\href {\doibase
  10.1088/1674-1056/ac3222} {\bibfield  {journal} {\bibinfo  {journal} {Chin.
  Phys. B}\ }\textbf {\bibinfo {volume} {31}},\ \bibinfo {pages} {017401}
  (\bibinfo {year} {2022})}\BibitemShut {NoStop}%
\bibitem [{\citenamefont {Li}\ \emph {et~al.}(2022)\citenamefont {Li},
  \citenamefont {Li}, \citenamefont {Gao},\ and\ \citenamefont
  {Tong}}]{PhysRevB.105.104201}%
  \BibitemOpen
  \bibfield  {author} {\bibinfo {author} {\bibfnamefont {S.}~\bibnamefont
  {Li}}, \bibinfo {author} {\bibfnamefont {M.}~\bibnamefont {Li}}, \bibinfo
  {author} {\bibfnamefont {Y.}~\bibnamefont {Gao}}, \ and\ \bibinfo {author}
  {\bibfnamefont {P.}~\bibnamefont {Tong}},\ }\bibfield  {title} {\enquote
  {\bibinfo {title} {Topological properties and localization transition in a
  one-dimensional non-hermitian lattice with a slowly varying potential},}\
  }\href {\doibase 10.1103/PhysRevB.105.104201} {\bibfield  {journal} {\bibinfo
   {journal} {Phys. Rev. B}\ }\textbf {\bibinfo {volume} {105}},\ \bibinfo
  {pages} {104201} (\bibinfo {year} {2022})}\BibitemShut {NoStop}%
\bibitem [{\citenamefont {Farchioni}\ \emph {et~al.}(1993)\citenamefont
  {Farchioni}, \citenamefont {Grosso},\ and\ \citenamefont
  {Parravicini}}]{Farchioni1993IncommensuratePA}%
  \BibitemOpen
  \bibfield  {author} {\bibinfo {author} {\bibfnamefont {R}~\bibnamefont
  {Farchioni}}, \bibinfo {author} {\bibfnamefont {G}~\bibnamefont {Grosso}}, \
  and\ \bibinfo {author} {\bibfnamefont {G~P}\ \bibnamefont {Parravicini}},\
  }\bibfield  {title} {\enquote {\bibinfo {title} {Incommensurate potentials:
  analytic and numerical progress},}\ }\href {\doibase
  10.1088/0953-8984/5/34B/003} {\bibfield  {journal} {\bibinfo  {journal} {J.
  Phys.: Condens. Matter}\ }\textbf {\bibinfo {volume} {5}},\ \bibinfo {pages}
  {B13} (\bibinfo {year} {1993})}\BibitemShut {NoStop}%
\bibitem [{\citenamefont {Li}\ and\ \citenamefont
  {Das~Sarma}(2020{\natexlab{a}})}]{art}%
  \BibitemOpen
  \bibfield  {author} {\bibinfo {author} {\bibfnamefont {X.}~\bibnamefont
  {Li}}\ and\ \bibinfo {author} {\bibfnamefont {S.}~\bibnamefont {Das~Sarma}},\
  }\bibfield  {title} {\enquote {\bibinfo {title} {Mobility edge and
  intermediate phase in one-dimensional incommensurate lattice potentials},}\
  }\href {\doibase 10.1103/PhysRevB.101.064203} {\bibfield  {journal} {\bibinfo
   {journal} {Phys. Rev. B}\ }\textbf {\bibinfo {volume} {101}},\ \bibinfo
  {pages} {064203} (\bibinfo {year} {2020}{\natexlab{a}})}\BibitemShut
  {NoStop}%
\bibitem [{\citenamefont {Yuce}(2014)}]{YUCE20142024}%
  \BibitemOpen
  \bibfield  {author} {\bibinfo {author} {\bibfnamefont {C.}~\bibnamefont
  {Yuce}},\ }\bibfield  {title} {\enquote {\bibinfo {title} {Pt symmetric
  aubry–andr\'{e} model},}\ }\href {\doibase
  https://doi.org/10.1016/j.physleta.2014.05.005} {\bibfield  {journal}
  {\bibinfo  {journal} {Phys. Lett. A}\ }\textbf {\bibinfo {volume} {378}},\
  \bibinfo {pages} {2024} (\bibinfo {year} {2014})}\BibitemShut {NoStop}%
\bibitem [{\citenamefont {Zeng}\ \emph {et~al.}(2017)\citenamefont {Zeng},
  \citenamefont {Chen},\ and\ \citenamefont {L\"u}}]{PhysRevA.95.062118}%
  \BibitemOpen
  \bibfield  {author} {\bibinfo {author} {\bibfnamefont {Q.-B.}\ \bibnamefont
  {Zeng}}, \bibinfo {author} {\bibfnamefont {S.}~\bibnamefont {Chen}}, \ and\
  \bibinfo {author} {\bibfnamefont {R.}~\bibnamefont {L\"u}},\ }\bibfield
  {title} {\enquote {\bibinfo {title} {Anderson localization in the
  non-hermitian aubry-andr\'e-harper model with physical gain and loss},}\
  }\href {\doibase 10.1103/PhysRevA.95.062118} {\bibfield  {journal} {\bibinfo
  {journal} {Phys. Rev. A}\ }\textbf {\bibinfo {volume} {95}},\ \bibinfo
  {pages} {062118} (\bibinfo {year} {2017})}\BibitemShut {NoStop}%
\bibitem [{\citenamefont {Jiang}\ \emph {et~al.}(2019)\citenamefont {Jiang},
  \citenamefont {Lang}, \citenamefont {Yang}, \citenamefont {Zhu},\ and\
  \citenamefont {Chen}}]{PhysRevB.100.054301}%
  \BibitemOpen
  \bibfield  {author} {\bibinfo {author} {\bibfnamefont {H.}~\bibnamefont
  {Jiang}}, \bibinfo {author} {\bibfnamefont {L.-J.}\ \bibnamefont {Lang}},
  \bibinfo {author} {\bibfnamefont {C.}~\bibnamefont {Yang}}, \bibinfo {author}
  {\bibfnamefont {S.-L.}\ \bibnamefont {Zhu}}, \ and\ \bibinfo {author}
  {\bibfnamefont {S.}~\bibnamefont {Chen}},\ }\bibfield  {title} {\enquote
  {\bibinfo {title} {Interplay of non-hermitian skin effects and anderson
  localization in nonreciprocal quasiperiodic lattices},}\ }\href {\doibase
  10.1103/PhysRevB.100.054301} {\bibfield  {journal} {\bibinfo  {journal}
  {Phys. Rev. B}\ }\textbf {\bibinfo {volume} {100}},\ \bibinfo {pages}
  {054301} (\bibinfo {year} {2019})}\BibitemShut {NoStop}%
\bibitem [{\citenamefont {Longhi}(2019)}]{PhysRevLett.122.237601}%
  \BibitemOpen
  \bibfield  {author} {\bibinfo {author} {\bibfnamefont {S.}~\bibnamefont
  {Longhi}},\ }\bibfield  {title} {\enquote {\bibinfo {title} {Topological
  phase transition in non-hermitian quasicrystals},}\ }\href {\doibase
  10.1103/PhysRevLett.122.237601} {\bibfield  {journal} {\bibinfo  {journal}
  {Phys. Rev. Lett.}\ }\textbf {\bibinfo {volume} {122}},\ \bibinfo {pages}
  {237601} (\bibinfo {year} {2019})}\BibitemShut {NoStop}%
\bibitem [{\citenamefont {Zeng}\ and\ \citenamefont
  {Xu}(2020)}]{PhysRevResearch.2.033052}%
  \BibitemOpen
  \bibfield  {author} {\bibinfo {author} {\bibfnamefont {Q.-B.}\ \bibnamefont
  {Zeng}}\ and\ \bibinfo {author} {\bibfnamefont {Y.}~\bibnamefont {Xu}},\
  }\bibfield  {title} {\enquote {\bibinfo {title} {Winding numbers and
  generalized mobility edges in non-hermitian systems},}\ }\href {\doibase
  10.1103/PhysRevResearch.2.033052} {\bibfield  {journal} {\bibinfo  {journal}
  {Phys. Rev. Res.}\ }\textbf {\bibinfo {volume} {2}},\ \bibinfo {pages}
  {033052} (\bibinfo {year} {2020})}\BibitemShut {NoStop}%
\bibitem [{\citenamefont {M\"uller}\ \emph {et~al.}(2016)\citenamefont
  {M\"uller}, \citenamefont {Delande},\ and\ \citenamefont
  {Shapiro}}]{PhysRevA.94.033615}%
  \BibitemOpen
  \bibfield  {author} {\bibinfo {author} {\bibfnamefont {C.~A.}\ \bibnamefont
  {M\"uller}}, \bibinfo {author} {\bibfnamefont {D.}~\bibnamefont {Delande}}, \
  and\ \bibinfo {author} {\bibfnamefont {B.}~\bibnamefont {Shapiro}},\
  }\bibfield  {title} {\enquote {\bibinfo {title} {Critical dynamics at the
  anderson localization mobility edge},}\ }\href {\doibase
  10.1103/PhysRevA.94.033615} {\bibfield  {journal} {\bibinfo  {journal} {Phys.
  Rev. A}\ }\textbf {\bibinfo {volume} {94}},\ \bibinfo {pages} {033615}
  (\bibinfo {year} {2016})}\BibitemShut {NoStop}%
\bibitem [{\citenamefont {Dai}\ \emph {et~al.}(2018)\citenamefont {Dai},
  \citenamefont {Wang},\ and\ \citenamefont {Yi}}]{PhysRevA.98.013635}%
  \BibitemOpen
  \bibfield  {author} {\bibinfo {author} {\bibfnamefont {C.~M.}\ \bibnamefont
  {Dai}}, \bibinfo {author} {\bibfnamefont {W.}~\bibnamefont {Wang}}, \ and\
  \bibinfo {author} {\bibfnamefont {X.~X.}\ \bibnamefont {Yi}},\ }\bibfield
  {title} {\enquote {\bibinfo {title} {Dynamical localization-delocalization
  crossover in the aubry-andr\'e-harper model},}\ }\href {\doibase
  10.1103/PhysRevA.98.013635} {\bibfield  {journal} {\bibinfo  {journal} {Phys.
  Rev. A}\ }\textbf {\bibinfo {volume} {98}},\ \bibinfo {pages} {013635}
  (\bibinfo {year} {2018})}\BibitemShut {NoStop}%
\bibitem [{\citenamefont {Qin}\ \emph {et~al.}(2014)\citenamefont {Qin},
  \citenamefont {Yin},\ and\ \citenamefont {Chen}}]{PhysRevB.90.054303}%
  \BibitemOpen
  \bibfield  {author} {\bibinfo {author} {\bibfnamefont {P.}~\bibnamefont
  {Qin}}, \bibinfo {author} {\bibfnamefont {C.}~\bibnamefont {Yin}}, \ and\
  \bibinfo {author} {\bibfnamefont {S.}~\bibnamefont {Chen}},\ }\bibfield
  {title} {\enquote {\bibinfo {title} {Dynamical anderson transition in
  one-dimensional periodically kicked incommensurate lattices},}\ }\href
  {\doibase 10.1103/PhysRevB.90.054303} {\bibfield  {journal} {\bibinfo
  {journal} {Phys. Rev. B}\ }\textbf {\bibinfo {volume} {90}},\ \bibinfo
  {pages} {054303} (\bibinfo {year} {2014})}\BibitemShut {NoStop}%
\bibitem [{\citenamefont {Nilanjan}\ and\ \citenamefont
  {Auditya}(2021)}]{Roy_2021}%
  \BibitemOpen
  \bibfield  {author} {\bibinfo {author} {\bibfnamefont {R.}~\bibnamefont
  {Nilanjan}}\ and\ \bibinfo {author} {\bibfnamefont {S.}~\bibnamefont
  {Auditya}},\ }\bibfield  {title} {\enquote {\bibinfo {title} {Entanglement
  entropy and out-of-time-order correlator in the long-range
  aubry–andré–harper model},}\ }\href {\doibase 10.1088/1361-648X/ac06e9}
  {\bibfield  {journal} {\bibinfo  {journal} {J. Phys.: Condens. Matter}\
  }\textbf {\bibinfo {volume} {33}},\ \bibinfo {pages} {334001} (\bibinfo
  {year} {2021})}\BibitemShut {NoStop}%
\bibitem [{\citenamefont {Li}\ and\ \citenamefont
  {Das~Sarma}(2020{\natexlab{b}})}]{PhysRevB.101.064203}%
  \BibitemOpen
  \bibfield  {author} {\bibinfo {author} {\bibfnamefont {X.}~\bibnamefont
  {Li}}\ and\ \bibinfo {author} {\bibfnamefont {S.}~\bibnamefont {Das~Sarma}},\
  }\bibfield  {title} {\enquote {\bibinfo {title} {Mobility edge and
  intermediate phase in one-dimensional incommensurate lattice potentials},}\
  }\href {\doibase 10.1103/PhysRevB.101.064203} {\bibfield  {journal} {\bibinfo
   {journal} {Phys. Rev. B}\ }\textbf {\bibinfo {volume} {101}},\ \bibinfo
  {pages} {064203} (\bibinfo {year} {2020}{\natexlab{b}})}\BibitemShut
  {NoStop}%
\bibitem [{\citenamefont {Li}\ \emph {et~al.}(2017)\citenamefont {Li},
  \citenamefont {Li},\ and\ \citenamefont {Das~Sarma}}]{PhysRevB.96.085119}%
  \BibitemOpen
  \bibfield  {author} {\bibinfo {author} {\bibfnamefont {X.}~\bibnamefont
  {Li}}, \bibinfo {author} {\bibfnamefont {X.}~\bibnamefont {Li}}, \ and\
  \bibinfo {author} {\bibfnamefont {S.}~\bibnamefont {Das~Sarma}},\ }\bibfield
  {title} {\enquote {\bibinfo {title} {Mobility edges in one-dimensional
  bichromatic incommensurate potentials},}\ }\href {\doibase
  10.1103/PhysRevB.96.085119} {\bibfield  {journal} {\bibinfo  {journal} {Phys.
  Rev. B}\ }\textbf {\bibinfo {volume} {96}},\ \bibinfo {pages} {085119}
  (\bibinfo {year} {2017})}\BibitemShut {NoStop}%
\bibitem [{\citenamefont {Weidemann}\ \emph {et~al.}(2022)\citenamefont
  {Weidemann}, \citenamefont {Kremer}, \citenamefont {Longhi},\ and\
  \citenamefont {Szameit}}]{weidemann2022topological}%
  \BibitemOpen
  \bibfield  {author} {\bibinfo {author} {\bibfnamefont {S.}~\bibnamefont
  {Weidemann}}, \bibinfo {author} {\bibfnamefont {M.}~\bibnamefont {Kremer}},
  \bibinfo {author} {\bibfnamefont {S.}~\bibnamefont {Longhi}}, \ and\ \bibinfo
  {author} {\bibfnamefont {A.}~\bibnamefont {Szameit}},\ }\bibfield  {title}
  {\enquote {\bibinfo {title} {Topological triple phase transition in
  non-hermitian floquet quasicrystals},}\ }\href
  {https://doi.org/10.1038/s41586-021-04253-0} {\bibfield  {journal} {\bibinfo
  {journal} {Nature}\ }\textbf {\bibinfo {volume} {601}},\ \bibinfo {pages}
  {354} (\bibinfo {year} {2022})}\BibitemShut {NoStop}%
\bibitem [{\citenamefont {Longhi}(2021)}]{PhysRevB.103.144202}%
  \BibitemOpen
  \bibfield  {author} {\bibinfo {author} {\bibfnamefont {S.}~\bibnamefont
  {Longhi}},\ }\bibfield  {title} {\enquote {\bibinfo {title} {Spectral
  deformations in non-hermitian lattices with disorder and skin effect: A
  solvable model},}\ }\href {\doibase 10.1103/PhysRevB.103.144202} {\bibfield
  {journal} {\bibinfo  {journal} {Phys. Rev. B}\ }\textbf {\bibinfo {volume}
  {103}},\ \bibinfo {pages} {144202} (\bibinfo {year} {2021})}\BibitemShut
  {NoStop}%
\bibitem [{\citenamefont {Tang}\ \emph {et~al.}(2022)\citenamefont {Tang},
  \citenamefont {Liu}, \citenamefont {Zhang},\ and\ \citenamefont
  {Zhang}}]{PhysRevA.105.063327}%
  \BibitemOpen
  \bibfield  {author} {\bibinfo {author} {\bibfnamefont {L.-Z.}\ \bibnamefont
  {Tang}}, \bibinfo {author} {\bibfnamefont {S.-N.}\ \bibnamefont {Liu}},
  \bibinfo {author} {\bibfnamefont {G.-Q.}\ \bibnamefont {Zhang}}, \ and\
  \bibinfo {author} {\bibfnamefont {D.-W.}\ \bibnamefont {Zhang}},\ }\bibfield
  {title} {\enquote {\bibinfo {title} {Topological anderson insulators with
  different bulk states in quasiperiodic chains},}\ }\href {\doibase
  10.1103/PhysRevA.105.063327} {\bibfield  {journal} {\bibinfo  {journal}
  {Phys. Rev. A}\ }\textbf {\bibinfo {volume} {105}},\ \bibinfo {pages}
  {063327} (\bibinfo {year} {2022})}\BibitemShut {NoStop}%
\bibitem [{\citenamefont {Mu}\ \emph {et~al.}(2022)\citenamefont {Mu},
  \citenamefont {Zhou}, \citenamefont {Li},\ and\ \citenamefont
  {Gong}}]{PhysRevB.105.205402}%
  \BibitemOpen
  \bibfield  {author} {\bibinfo {author} {\bibfnamefont {S.}~\bibnamefont
  {Mu}}, \bibinfo {author} {\bibfnamefont {L.}~\bibnamefont {Zhou}}, \bibinfo
  {author} {\bibfnamefont {L.}~\bibnamefont {Li}}, \ and\ \bibinfo {author}
  {\bibfnamefont {J.}~\bibnamefont {Gong}},\ }\bibfield  {title} {\enquote
  {\bibinfo {title} {Non-hermitian pseudo mobility edge in a coupled chain
  system},}\ }\href {\doibase 10.1103/PhysRevB.105.205402} {\bibfield
  {journal} {\bibinfo  {journal} {Phys. Rev. B}\ }\textbf {\bibinfo {volume}
  {105}},\ \bibinfo {pages} {205402} (\bibinfo {year} {2022})}\BibitemShut
  {NoStop}%
\bibitem [{\citenamefont {Huang}\ \emph {et~al.}()\citenamefont {Huang},
  \citenamefont {Vu}, \citenamefont {Li},\ and\ \citenamefont
  {Sarma}}]{unknown}%
  \BibitemOpen
  \bibfield  {author} {\bibinfo {author} {\bibfnamefont {K.}~\bibnamefont
  {Huang}}, \bibinfo {author} {\bibfnamefont {D.}~\bibnamefont {Vu}}, \bibinfo
  {author} {\bibfnamefont {X.}~\bibnamefont {Li}}, \ and\ \bibinfo {author}
  {\bibfnamefont {S.}~\bibnamefont {Sarma}},\ }\bibfield  {title} {\enquote
  {\bibinfo {title} {Incommensurate many-body localization in the presence of
  long-range hopping and single-particle mobility edge},}\ }\href {\doibase
  arXiv:2205.15339} {\ arXiv:2205.15339}\BibitemShut {NoStop}%
\bibitem [{\citenamefont {Pranjal}\ \emph {et~al.}(2017)\citenamefont
  {Pranjal}, \citenamefont {Henrik}, \citenamefont {Ulrich}, \citenamefont
  {Michael},\ and\ \citenamefont {Immanuel}}]{Bordia_2017}%
  \BibitemOpen
  \bibfield  {author} {\bibinfo {author} {\bibfnamefont {B.}~\bibnamefont
  {Pranjal}}, \bibinfo {author} {\bibfnamefont {L.}~\bibnamefont {Henrik}},
  \bibinfo {author} {\bibfnamefont {S.}~\bibnamefont {Ulrich}}, \bibinfo
  {author} {\bibfnamefont {K.}~\bibnamefont {Michael}}, \ and\ \bibinfo
  {author} {\bibfnamefont {B.}~\bibnamefont {Immanuel}},\ }\bibfield  {title}
  {\enquote {\bibinfo {title} {Periodically driving a many-body localized
  quantum system},}\ }\href {\doibase 10.1038/nphys4020} {\bibfield  {journal}
  {\bibinfo  {journal} {Nat. Phys.}\ }\textbf {\bibinfo {volume} {13}},\
  \bibinfo {pages} {460} (\bibinfo {year} {2017})}\BibitemShut {NoStop}%
\bibitem [{\citenamefont {Hamazaki}\ \emph {et~al.}(2019)\citenamefont
  {Hamazaki}, \citenamefont {Kawabata},\ and\ \citenamefont
  {Ueda}}]{PhysRevLett.123.090603}%
  \BibitemOpen
  \bibfield  {author} {\bibinfo {author} {\bibfnamefont {R.}~\bibnamefont
  {Hamazaki}}, \bibinfo {author} {\bibfnamefont {K.}~\bibnamefont {Kawabata}},
  \ and\ \bibinfo {author} {\bibfnamefont {M.}~\bibnamefont {Ueda}},\
  }\bibfield  {title} {\enquote {\bibinfo {title} {Non-hermitian many-body
  localization},}\ }\href {\doibase 10.1103/PhysRevLett.123.090603} {\bibfield
  {journal} {\bibinfo  {journal} {Phys. Rev. Lett.}\ }\textbf {\bibinfo
  {volume} {123}},\ \bibinfo {pages} {090603} (\bibinfo {year}
  {2019})}\BibitemShut {NoStop}%
\bibitem [{\citenamefont {L\"uschen}\ \emph {et~al.}(2018)\citenamefont
  {L\"uschen}, \citenamefont {Scherg}, \citenamefont {Kohlert}, \citenamefont
  {Schreiber}, \citenamefont {Bordia}, \citenamefont {Li}, \citenamefont
  {Das~Sarma},\ and\ \citenamefont {Bloch}}]{PhysRevLett.120.160404}%
  \BibitemOpen
  \bibfield  {author} {\bibinfo {author} {\bibfnamefont {H.~P.}\ \bibnamefont
  {L\"uschen}}, \bibinfo {author} {\bibfnamefont {S.}~\bibnamefont {Scherg}},
  \bibinfo {author} {\bibfnamefont {T.}~\bibnamefont {Kohlert}}, \bibinfo
  {author} {\bibfnamefont {M.}~\bibnamefont {Schreiber}}, \bibinfo {author}
  {\bibfnamefont {P.}~\bibnamefont {Bordia}}, \bibinfo {author} {\bibfnamefont
  {X.}~\bibnamefont {Li}}, \bibinfo {author} {\bibfnamefont {S.}~\bibnamefont
  {Das~Sarma}}, \ and\ \bibinfo {author} {\bibfnamefont {I.}~\bibnamefont
  {Bloch}},\ }\bibfield  {title} {\enquote {\bibinfo {title} {Single-particle
  mobility edge in a one-dimensional quasiperiodic optical lattice},}\ }\href
  {\doibase 10.1103/PhysRevLett.120.160404} {\bibfield  {journal} {\bibinfo
  {journal} {Phys. Rev. Lett.}\ }\textbf {\bibinfo {volume} {120}},\ \bibinfo
  {pages} {160404} (\bibinfo {year} {2018})}\BibitemShut {NoStop}%
\bibitem [{\citenamefont {Xu}\ \emph {et~al.}(2021)\citenamefont {Xu},
  \citenamefont {Xia},\ and\ \citenamefont {Chen}}]{PhysRevB.104.224204}%
  \BibitemOpen
  \bibfield  {author} {\bibinfo {author} {\bibfnamefont {Z.}~\bibnamefont
  {Xu}}, \bibinfo {author} {\bibfnamefont {X.}~\bibnamefont {Xia}}, \ and\
  \bibinfo {author} {\bibfnamefont {S.}~\bibnamefont {Chen}},\ }\bibfield
  {title} {\enquote {\bibinfo {title} {Non-hermitian aubry-andr\'e model with
  power-law hopping},}\ }\href {\doibase 10.1103/PhysRevB.104.224204}
  {\bibfield  {journal} {\bibinfo  {journal} {Phys. Rev. B}\ }\textbf {\bibinfo
  {volume} {104}},\ \bibinfo {pages} {224204} (\bibinfo {year}
  {2021})}\BibitemShut {NoStop}%
\bibitem [{\citenamefont {Chhabra}\ and\ \citenamefont
  {Jensen}(1989)}]{PhysRevLett.62.1327}%
  \BibitemOpen
  \bibfield  {author} {\bibinfo {author} {\bibfnamefont {A.}~\bibnamefont
  {Chhabra}}\ and\ \bibinfo {author} {\bibfnamefont {R.~V.}\ \bibnamefont
  {Jensen}},\ }\bibfield  {title} {\enquote {\bibinfo {title} {Direct
  determination of the f(\ensuremath{\alpha}) singularity spectrum},}\ }\href
  {\doibase 10.1103/PhysRevLett.62.1327} {\bibfield  {journal} {\bibinfo
  {journal} {Phys. Rev. Lett.}\ }\textbf {\bibinfo {volume} {62}},\ \bibinfo
  {pages} {1327} (\bibinfo {year} {1989})}\BibitemShut {NoStop}%
\bibitem [{\citenamefont {Janssen}(1994)}]{doi:10.1142/S021797929400049X}%
  \BibitemOpen
  \bibfield  {author} {\bibinfo {author} {\bibfnamefont {M.}~\bibnamefont
  {Janssen}},\ }\bibfield  {title} {\enquote {\bibinfo {title} {Multifractal
  analysis of broadly-distributed observables at criticality},}\ }\href
  {\doibase 10.1142/S021797929400049X} {\bibfield  {journal} {\bibinfo
  {journal} {Int. J. Mod. Phys. B}\ }\textbf {\bibinfo {volume} {08}},\
  \bibinfo {pages} {943} (\bibinfo {year} {1994})}\BibitemShut {NoStop}%
\bibitem [{\citenamefont {Huckestein}(1995)}]{RevModPhys.67.357}%
  \BibitemOpen
  \bibfield  {author} {\bibinfo {author} {\bibfnamefont {B.}~\bibnamefont
  {Huckestein}},\ }\bibfield  {title} {\enquote {\bibinfo {title} {Scaling
  theory of the integer quantum hall effect},}\ }\href {\doibase
  10.1103/RevModPhys.67.357} {\bibfield  {journal} {\bibinfo  {journal} {Rev.
  Mod. Phys.}\ }\textbf {\bibinfo {volume} {67}},\ \bibinfo {pages} {357}
  (\bibinfo {year} {1995})}\BibitemShut {NoStop}%
\bibitem [{\citenamefont {Cuevas}(2003)}]{PhysRevB.68.184206}%
  \BibitemOpen
  \bibfield  {author} {\bibinfo {author} {\bibfnamefont {E.}~\bibnamefont
  {Cuevas}},\ }\bibfield  {title} {\enquote {\bibinfo {title} {Multifractality
  of hamiltonians with power-law transfer terms},}\ }\href {\doibase
  10.1103/PhysRevB.68.184206} {\bibfield  {journal} {\bibinfo  {journal} {Phys.
  Rev. B}\ }\textbf {\bibinfo {volume} {68}},\ \bibinfo {pages} {184206}
  (\bibinfo {year} {2003})}\BibitemShut {NoStop}%
\bibitem [{\citenamefont {Simon}\ and\ \citenamefont
  {Spencer}(1989)}]{simon1989trace}%
  \BibitemOpen
  \bibfield  {author} {\bibinfo {author} {\bibfnamefont {B.}~\bibnamefont
  {Simon}}\ and\ \bibinfo {author} {\bibfnamefont {T.}~\bibnamefont
  {Spencer}},\ }\bibfield  {title} {\enquote {\bibinfo {title} {Trace class
  perturbations and the absence of absolutely continuous spectra},}\ }\href
  {https://doi.org/10.1007/BF01217772} {\bibfield  {journal} {\bibinfo
  {journal} {Commun. Math. Phys.}\ }\textbf {\bibinfo {volume} {125}},\
  \bibinfo {pages} {113} (\bibinfo {year} {1989})}\BibitemShut {NoStop}%
\bibitem [{\citenamefont {Simon}(2005)}]{simon2005trace}%
  \BibitemOpen
  \bibfield  {author} {\bibinfo {author} {\bibfnamefont {B.}~\bibnamefont
  {Simon}},\ }\href@noop {} {\emph {\bibinfo {title} {Trace ideals and their
  applications}}},\ \bibinfo {number} {120}\ (\bibinfo  {publisher} {American
  Mathematical Soc.},\ \bibinfo {year} {2005})\BibitemShut {NoStop}%
\bibitem [{\citenamefont {Avila}(2015)}]{avila2015global}%
  \BibitemOpen
  \bibfield  {author} {\bibinfo {author} {\bibfnamefont {A.}~\bibnamefont
  {Avila}},\ }\bibfield  {title} {\enquote {\bibinfo {title} {Global theory of
  one-frequency schr{\"o}dinger operators},}\ }\href {\doibase
  10.1007/s11511-015-0128-7} {\bibfield  {journal} {\bibinfo  {journal} {Acta
  Math.}\ }\textbf {\bibinfo {volume} {215}},\ \bibinfo {pages} {1} (\bibinfo
  {year} {2015})}\BibitemShut {NoStop}%
\bibitem [{\citenamefont {Cai}\ and\ \citenamefont {Yu}(2022)}]{cai}%
  \BibitemOpen
  \bibfield  {author} {\bibinfo {author} {\bibfnamefont {X.}~\bibnamefont
  {Cai}}\ and\ \bibinfo {author} {\bibfnamefont {Y.-C.}\ \bibnamefont {Yu}},\
  }\bibfield  {title} {\enquote {\bibinfo {title} {Exact mobility edges in
  quasiperiodic systems without self-duality},}\ }\href {\doibase
  10.1088/1361-648X/aca136} {\bibfield  {journal} {\bibinfo  {journal} {J.
  Phys.: Condens. Matter}\ }\textbf {\bibinfo {volume} {35}},\ \bibinfo {pages}
  {035602} (\bibinfo {year} {2022})}\BibitemShut {NoStop}%
\bibitem [{\citenamefont {Sarnak}(1982)}]{sarnak1982spectral}%
  \BibitemOpen
  \bibfield  {author} {\bibinfo {author} {\bibfnamefont {P.}~\bibnamefont
  {Sarnak}},\ }\bibfield  {title} {\enquote {\bibinfo {title} {Spectral
  behavior of quasi periodic potentials},}\ }\href
  {https://doi.org/10.1007/BF01208483} {\bibfield  {journal} {\bibinfo
  {journal} {Commun. Math. Phys.}\ }\textbf {\bibinfo {volume} {84}},\ \bibinfo
  {pages} {377} (\bibinfo {year} {1982})}\BibitemShut {NoStop}%
\bibitem [{\citenamefont {Xu}\ \emph {et~al.}(2022)\citenamefont {Xu},
  \citenamefont {Xia},\ and\ \citenamefont {Chen}}]{xu2022exact}%
  \BibitemOpen
  \bibfield  {author} {\bibinfo {author} {\bibfnamefont {Z.}~\bibnamefont
  {Xu}}, \bibinfo {author} {\bibfnamefont {X.}~\bibnamefont {Xia}}, \ and\
  \bibinfo {author} {\bibfnamefont {S.}~\bibnamefont {Chen}},\ }\bibfield
  {title} {\enquote {\bibinfo {title} {Exact mobility edges and topological
  phase transition in two-dimensional non-hermitian quasicrystals},}\ }\href
  {https://doi.org/10.1007/s11433-021-1802-4} {\bibfield  {journal} {\bibinfo
  {journal} {Sci. China Phys. Mech. Astron.}\ }\textbf {\bibinfo {volume}
  {65}},\ \bibinfo {pages} {1} (\bibinfo {year} {2022})}\BibitemShut {NoStop}%
\end{thebibliography}%
\end{document}